\documentclass[
    aps,
    prd,
    superscriptaddress,
    twocolumn,
    a4paper,
    floatfix,
    bibliography,
    nofootinbib,
    preprintnumbers
]{revtex4-2}

\usepackage[svgnames]{xcolor}

\usepackage[normalem]{ulem}

\usepackage[american]{babel}
\usepackage[utf8x]{inputenc}

\usepackage{grffile} 

\usepackage{newtxtext,newtxmath}
\usepackage{microtype}

\usepackage{amsmath}
\usepackage{bbm}
\usepackage{bm}
\usepackage{mathtools}
\usepackage{dsfont}
\usepackage{braket}
\usepackage{cancel}
\usepackage{slashed}

\usepackage{graphicx}

\usepackage{siunitx} 

\usepackage{ragged2e}
\usepackage{array}
\usepackage{tabularx}
\usepackage{booktabs}
\usepackage{makecell}

\definecolor{cset-aps-blueberry}{RGB}{28,128,158}
\definecolor{cset-aps-blue}{RGB}{46,44,184}
\definecolor{cset-aps-turquoise}{RGB}{0,67,88}
\definecolor{cset-aps-limegreen}{RGB}{190,219,67}
\definecolor{cset-aps-green}{RGB}{31,138,112}
\definecolor{cset-aps-yellow}{RGB}{255,225,25}
\definecolor{cset-aps-orange}{RGB}{253,116,0}
\definecolor{cset-aps-red}{RGB}{219,0,43}

\usepackage{tikz}

\usepackage{pgfplots}
\pgfplotsset{%
    every axis legend/.append style={%
        cells={anchor=west},
        at={(0.96,0.04)},
        anchor=south east,
        font=\scriptsize,
        },
    every axis/.append style={%
        yticklabel style={%
            /pgf/number format/fixed zerofill,
            /pgf/number format/precision=2},
        },
    width= \textwidth,
    height=8cm,
    xmajorgrids=true,
    xminorgrids=false,
    minor x tick num=1,
}
\usepgfplotslibrary{external}

\usetikzlibrary{decorations}

\usepackage{pict2e,picture}

\makeatletter
\DeclareRobustCommand{\Arrow}[1][]{%
\check@mathfonts
\if\relax\detokenize{#1}\relax
\settowidth{\dimen@}{$\m@th\rightarrow$}%
\else
\setlength{\dimen@}{#1}%
\fi
\sbox\z@{\usefont{U}{lasy}{m}{n}\symbol{41}}%
\begin{picture}(\dimen@,\ht\z@)
\roundcap
\put(\dimexpr\dimen@-.7\wd\z@,0){\usebox\z@}
\put(0,\fontdimen22\textfont2){\line(1,0){\dimen@}}
\end{picture}%
}
\makeatother

\usepackage{hyperref}
\hypersetup{%
    colorlinks=true,
    linkcolor={cset-aps-red},
    linkbordercolor={cset-aps-red},
    filecolor={cset-aps-orange},
    filebordercolor={cset-aps-orange},
    citecolor={cset-aps-blue},
    citebordercolor={cset-aps-blue},
    urlcolor={cset-aps-green},
    urlbordercolor={cset-aps-green},
    menucolor={cset-aps-limegreen},
    menubordercolor={cset-aps-limegreen},
    breaklinks=true,
    pdfborderstyle={/S/U/W 2},
    pdfpagemode=UseOutlines,
    pdfstartpage={1},
}

\newcommand{\ii}{\text{i}}

\newcommand{\vect}[1]{\boldsymbol{#1}}

\newcommand{\eg}{e.\,g., }
\newcommand{\ie}{i.\,e., }

\usepackage{mathtools}
\usepackage{newtxmath}

\newsavebox{\symbox}

\newcommand{\lrpartial}[0]{%
  \sbox{\symbox}{$\partial$}%
  \partial%
  \kern-1.1\wd\symbox%
  \overset{\leftrightarrow}{\footnotesize\vphantom{i}}%
}

\newcommand{\lrD}[0]{%
  \sbox{\symbox}{$\mathcal{D}$}%
  \mathcal{D}%
  \kern-0.9\wd\symbox%
  \overset{\leftrightarrow}{\footnotesize\vphantom{i}}%
}

\newcommand{\rD}[0]{%
  \sbox{\symbox}{$\mathcal{D}$}%
  \mathcal{D}%
  \kern-0.9\wd\symbox%
  \overset{\rightarrow}{\footnotesize\vphantom{i}}%
}

\newcommand{\lD}[0]{%
  \sbox{\symbox}{$\mathcal{D}$}%
  \mathcal{D}%
  \kern-0.9\wd\symbox%
  \overset{\leftarrow}{\footnotesize\vphantom{i}}%
}

\newcommand{\lrDalt}[0]{%
  \sbox{\symbox}{$\mathfrak{D}$}%
  \mathfrak{D}%
  \kern-0.9\wd\symbox%
  \overset{\leftrightarrow}{\footnotesize\vphantom{i}}%
}

\newcommand{\rDalt}[0]{%
  \sbox{\symbox}{$\mathfrak{D}$}%
  \mathfrak{D}%
  \kern-0.9\wd\symbox%
  \overset{\rightarrow}{\footnotesize\vphantom{i}}%
}

\newcommand{\lDalt}[0]{%
  \sbox{\symbox}{$\mathfrak{D}$}%
  \mathfrak{D}%
  \kern-0.9\wd\symbox%
  \overset{\leftarrow}{\footnotesize\vphantom{i}}%
}

\newcommand{\lrE}[0]{%
  \sbox{\symbox}{$\mathcal{E}$}%
  \mathcal{E}%
  \kern-0.9\wd\symbox%
  \overset{\leftrightarrow}{\footnotesize\vphantom{i}}%
}

\newcommand{\lrO}[0]{%
  \sbox{\symbox}{$\mathcal{O}$}%
  \mathcal{O}%
  \kern-0.9\wd\symbox%
  \overset{\leftrightarrow}{\footnotesize\vphantom{i}}%
}

\newcommand{\lO}[0]{%
  \sbox{\symbox}{$\mathcal{O}$}%
  \mathcal{O}%
  \kern-0.9\wd\symbox%
  \overset{\leftarrow}{\footnotesize\vphantom{i}}%
}

\newcommand{\rO}[0]{%
  \sbox{\symbox}{$\mathcal{O}$}%
  \mathcal{O}%
  \kern-0.9\wd\symbox%
  \overset{\rightarrow}{\footnotesize\vphantom{i}}%
}

\usepackage{wasysym}

\usepackage{diagbox}

\usepackage{simplewick}

\newcommand{\orcid}[1]{\href{https://orcid.org/#1}{\includegraphics[width=7pt]{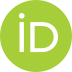}}}

\newcommand\blfootnote[1]{
    \begingroup
    \renewcommand\thefootnote{}\footnote{#1}
    \addtocounter{footnote}{-1}
    \endgroup
}

\newcommand{\sMin}{standard-Minkowski }

\begin{document}

\title{Fermion quantum field theory on curved and non-inertial backgrounds in \sMin form}

\newcommand{\affULM}{\address{Institut f{\"u}r Quantenphysik and Center for Integrated Quantum Science and Technology (IQST), Universit{\"a}t Ulm, Albert-Einstein-Allee 11, D-89081 Ulm, Germany}}
\newcommand{\affTUDa}{\affiliation{Technische Universit{\"a}t Darmstadt, Fachbereich Physik, Institut f{\"u}r Angewandte Physik, Schlossgartenstr. 7, D-64289 Darmstadt, Germany}}
\newcommand{\affYNU}{\affiliation{Department of Physics, Graduate School of Engineering Science, Yokohama National University, 79-5 Tokiwadai, Hodogaya-ku, Yokohama 240-8501, Japan}}
\newcommand{\affWAY}{\affiliation{Department of Physics and Astronomy, Wayne State University, Detroit, Michigan 48201, USA}}
\newcommand{\affDLR}{\affiliation{German Aerospace Center (DLR), Institute of Quantum Technologies, Wilhelm-Runge-Straße 10, D-89081 Ulm, Germany}}

\author{Fabio Di Pumpo$^{\ddag}$\orcid{0000-0002-6304-6183}}
\email{fabio.di-pumpo@uni-ulm.de}
\email{fabio.di-pumpo@gmx.de}
\affULM
\author{Tobias Asano$^{\ddag}$\orcid{0000-0002-6257-8815}}
\affYNU
\author{Alexander Friedrich\,\orcid{0000-0003-0588-1989}}\altaffiliation{Currently at the German Aerospace Center (DLR), Institute of Quantum Technologies, Wilhelm-Runge-Straße 10, D-89081 Ulm, Germany and ClockWerQ GmbH, Römerstraße 118, D-89077 Ulm, Germany.}
\affULM
\author{Gil Paz\,\orcid{0000-0001-5605-2043}}
\affWAY
\author{Enno Giese\,\orcid{0000-0002-1126-6352}\,}
\affTUDa

\preprint{WSU-HEP-2601} 

\begin{abstract}
\noindent
Quantum field theory on curved and non-inertial backgrounds contains background- and foliation-dependent quantities in the canonical Lagrangian, the hypersurface inner product and bilinear form, as well as in the equal-time anti-commutation relations.
In this work, we determine a local fermion-field redefinition that brings these canonical structures into their \sMin forms, \ie{} the forms they assume in Cartesian inertial coordinates on Minkowski spacetime, where the zeroth world coordinate is identified as the coordinate of time.
Starting from the generally covariant Dirac action minimally coupled to a spin-$1$ gauge field, we derive the corresponding Lagrangian, fermionic inner product, and quantization rule in an Arnowitt--Deser--Misner decomposition, formulated in arbitrary world coordinates.
We identify the generalized temporal gamma matrix as the common geometric factor governing the canonical temporal structure of all three quantities.
Using a field redefinition, we transform this generalized temporal gamma matrix to its \sMin form, thereby mapping the fermionic inner product and the equal-time anti-commutation relation to their \sMin expressions, while transferring the explicit background and foliation dependence to the transformed Lagrangian and fermion-field operators.
We show that such a field redefinition necessarily consists of a local rescaling and a fixing of the local Lorentz frame.
This procedure restores the conventional canonical normalization from \sMin spacetime used for fermionic mode quantization and occupation-number operators.
The transformed Lagrangian consequently assumes a generalized first-order Schrödinger form, leading to the familiar rest-energy term and spacetime-magnetic couplings, as well as to the leading non-relativistic limit, in which temporal derivatives are separated from spatial ones.
\end{abstract}

\maketitle
\section{Introduction}
\label{sec:intro}
\blfootnote{\hspace{-0.7em}${}^{\ddag}$ These authors contributed equally.}
\hspace{-1.5em} Many physical situations or experiments involve quantum fields propagating on curved, \ie{} gravitational, or non-inertial backgrounds.
However, gravity in its formulation through general relativity~\cite{Einstein1907,Weinberg1972,Misner1973,Wald1984,Will2014} is described as a classical geometrical theory of spacetime and the associated metric tensor.
A key feature of general relativity, compared to its inertial limit in special relativity~\cite{Einstein1905,Rindler1969} on a flat Minkowski spacetime, is the presence of intricate spacetime and observer dependence, encoded in the geometry and the choice of background.
Many physical quantities acquire a local and observer-dependent description, reflecting the absence of a preferred global inertial frame in a generic curved or non-inertial spacetime, although locally inertial frames exist at every point.
In contrast, special relativity, formulated on Minkowski spacetime, is based on fixed global geometric structures in inertial reference frames.
Quantum physics, in its modern formulation of quantum field theory~\cite{Peskin1995,Kleinert2016}, assigns quantized excitations to matter and gauge fields, providing a description of multi-particle and scattering processes.
Unlike general relativity, the non-gravitational interactions in conventional quantum field theory on Minkowski spacetime are described by dynamical fields propagating on the Minkowski background, rather than as part of the spacetime geometry itself.
As a result, a consistent framework unifying quantum field theory with gravity remains an open challenge~\cite{Kiefer2006}.

All known non-gravitational fundamental interactions are described by the Standard Model of particle physics~\cite{Gaillard1999,Thomson2013}, a quantum field theory for fundamental spin-$1/2$ fermions and a spin-$0$ Higgs boson, interacting via spin-$1$ gauge fields.
The Standard Model is typically formulated in Cartesian inertial coordinates on Minkowski spacetime, where fermionic fields and their excitations are often used to probe~\cite{Charpak1991,Moser2009,Malberti2026} fundamental interactions.
In this context, a formulation on Minkowski spacetime, more specifically an inertial and Cartesian one, assumes that the theory builds on global Poincar\'e symmetry~\cite{Poincare1906,Wigner1939,Weinberg1995} and is invariant under inertial, global Lorentz transformations~\cite{Einstein1905,Lorentz1937}.
These requirements define a Minkowski-type canonical framework for computing and measuring observables, as well as the quantization rule of the fields, in which all observer dependence is limited to an inertial and global special-relativistic notion.
These special-relativistic structures are highlighted through the Hermitian formulation~\cite{Weinberg1995} of the \sMin inner product $\left(\psi_1,\psi_2\right)=\int{\mathrm{d}^3x\,\psi^\dagger_1\psi_2}$ between two fermionic spinors $\psi_1$ and $\psi_2$.
In this context and in the following, we call objects and structures to be of \sMin form, if they assume a form as if the background were Minkowski spacetime and as if the zeroth component $\mu=\bar{0}$, defined below, of four world indices $\mu$ represented the spacetime-independent foliation-time function $x^{\bar{0}}=ct$, while the other three components encode spatial Cartesian indices.
Consequently, all relevant structures of textbook quantum field theories~\cite{Peskin1995,Kleinert2016} on Minkowski spacetime and with a static-time choice, \eg{} the Standard Model~\cite{Gaillard1999,Thomson2013}, are naturally given in this form.
This inner product provides the \sMin bilinear form~\cite{Streater1989,Peskin1995,Weinberg1995,Kleinert2016} for expectation values and transition elements, in accordance with the Born rule~\cite{Donald1992,Fewster2020,Papageorgiou2024} of quantum physics.
After a mode decomposition and vacuum have been chosen, this normalization also yields the conventional fermionic creation, annihilation, and occupation-number operators.
This bilinear form is connected to the spinors' inertial canonical quantization rule through the \sMin anti-commutator $\big\lbrace\psi_a\left(t,\vect{x}\right),\psi_b^\dagger\left(t,\vect{x}^\prime\right)\big\rbrace=\delta_{ab}\delta^{(3)}\left(\vect{x}-\vect{x}^\prime\right)$, with three-dimensional Dirac-delta distribution $\delta^{(3)}\left(\vect{x}-\vect{x}^\prime\right)$. 
These are the familiar coordinate expressions of the inner product and canonical quantization rule in Cartesian inertial coordinates on Minkowski spacetime, including the inertial structure and global Lorentz invariance of special relativity.
For a fermion with mass $m$, the \sMin Lagrangian density is $\mathcal{L}=\bar{\psi}\big[\ii\hbar c\delta^\mu_I\gamma^I \,\lrpartial_\mu/2-mc^2\big]\psi$, with inertial Minkowski gamma matrices $\gamma^I$ with local indices $I=0,1,2,3$, Kronecker delta $\delta^\mu_I$, speed of light $c$, symmetrized derivative $\bar{\psi}\delta^\mu_I\gamma^I \,\lrpartial_\mu\psi=\bar{\psi}\delta^\mu_I\gamma^I\left(\partial_\mu\psi\right)-\left(\partial_\mu\bar{\psi}\right)\delta^\mu_I\gamma^I\psi$, and Dirac adjoint $\bar{\psi}=\psi^\dagger\gamma^0$ relevant for the inertial Lorentz invariance. 
This Lagrangian has \sMin Schrödinger form $\mathcal{L}=\ii\hbar\psi^\dagger\lrpartial_t\psi/2-\mathcal{H}$, with Hamilton density $\mathcal{H}$, \ie{} apart from natural constants we find a unit prefactor in front of the zeroth-component derivative when using $\psi^\dagger$ instead of $\bar{\psi}$.
This prefactor is directly connected to the Minkowski background and the invariance under global Lorentz transformations, which is also retained when calculating the Dirac equation as the corresponding equation of motion through a variation~\cite{Peskin1995,Weinberg1995} of the action $S=\int{\mathrm{d}^4x\mathcal{L}}$, where the spacetime-volume element in Cartesian inertial coordinates is unity.
These \sMin properties provide the standard normalization used in mode decompositions and in defining fermionic excitations, which enter observables, \eg{} through the counting of fermionic excitations in scattering processes~\cite{Charpak1991,Moser2009,Malberti2026}.

However, quantum field theory on curved and non-inertial backgrounds, including gravity through curved spacetime~\cite{Misner1973} as well as other non-inertial effects~\cite{Ni1978} like rotations~\cite{Kajari2009} and accelerations~\cite{Rindler1969}, or to some extent even curvilinear coordinates, has been formulated~\cite{DeWitt1975,Birrell1982,Parker2009,Hollands2015,Buchbinder2021} based on invariance under world coordinate transformations (diffeomorphisms) and a local version of Lorentz transformations [local Lorentz transformations (LLTs)], in accordance with the geometric principles of general relativity.
Such theories have been used, \eg{} to describe the Unruh~\cite{Fulling1973,Davies1975,Unruh1976} and Hawking~\cite{Hawking1974,Wald2001} effects, leading to phenomena such as observer- and background-dependent particle notions~\cite{Crispino2008}, which do not arise from global Lorentz transformations in Minkowski quantum field theory and highlight the observer- and background-dependence of general relativity.
Thus, when formulating these quantum field theories on curved and non-inertial backgrounds, the explicit representation of structures connected to computations and measurements is modified compared to their \sMin forms.
As such, the inner product (or the fermion bilinear form) and the quantization rule at first no longer take~\cite{Leclerc2006,Leclerc2007,Egorov2024} their \sMin forms, rendering their interpretation and use less intuitive.
At the same time, also the naive Schrödinger form of the Lagrangian is no longer preserved, obscuring the intuitive foliation-time evolution of the dynamics.
As a consequence of these modifications, in the original field variables, curved and non-inertial backgrounds cannot simply be interpreted as an added force or interaction generating new terms, like in other modifications~\cite{Isidori2024} of the Standard Model.
Instead, non-inertial backgrounds lead to a more involved structure in these theories, which separates into two interconnected components:
the inner product (through the bilinear form) and the quantization rule on the one hand, and the dynamics encoded in the Lagrangian on the other.
Both components are in general affected by the curved and non-inertial background.

Our goal is a reformulation of quantum field theory for fermions on curved and non-inertial backgrounds such that for their measurement- and computation-related structures, namely the inner product and quantization rule, \sMin form is restored.
To this end, we introduce a field redefinition, consisting of a rescaling of spinors and a specific choice of the local Lorentz frame.
This reformulation naturally preserves the underlying symmetries under diffeomorphisms and LLTs, although covariance and the full LLT freedom are no longer manifest after choosing a foliation and a foliation-adapted local Lorentz frame.
While the full theory is coupled to a spin-$1$ gauge field and is, in principle, also affected by the non-inertial background, we focus in this work on the fermionic sector and the corresponding observables, as fermionic degrees of freedom are particularly relevant for measurements and signal computations~\cite{Charpak1991,Moser2009,Malberti2026}.
Moreover, this procedure removes the explicit background- and foliation-dependent factors from the inner product and quantization rule, while retaining the corresponding dependence in the transformed Lagrangian and in the relation between the original and redefined fields.
We also find that the redefinition restores the Lagrangian to a generalized Schrödinger form, and discuss the connection to the redefined inner product through the generalized temporal gamma matrix.
This way, we obtain a representation where gravity and non-inertial forces can indeed be interpreted, to a large extent, as additional interactions, while the \sMin interpretation of the Born-rule form and the usual bilinear bra-ket-like form of the inner product, together with the corresponding quantization rule, are recovered.

Our work is structured as follows:
In Sec.~\ref{sec:FieldTheoryBefore} we present a Lagrangian for the quantum field theory of a spin-$1/2$ field coupled to a spin-$1$ gauge field on curved and non-inertial backgrounds.
We introduce spacetime-dependent world-coordinate gamma matrices that encode also the spin and modified minimally-coupled derivatives.
In addition, we discuss the invariance of this theory under diffeomorphisms and LLTs.
Moreover, we introduce an Arnowitt--Deser--Misner (ADM) temporal and spatial decomposition~\cite{Arnowitt1960,Arnowitt2008} of the non-inertial background in arbitrary world coordinates based on a smooth (sufficiently differentiable
) spacelike foliation, and provide illustrative examples.
The spatially projected vierbein of the ADM decomposition defines a dreibein on each hypersurface, providing one of the geometric ingredients entering the Ashtekar formulation~\cite{Ashtekar1986}.
Then, we express the Lagrangian in this decomposition and analyze the implication of such a hypersurface on the non-inertial inner product, the bilinear spinor form, and the non-inertial quantization rule for the fermion field.
In this context, we identify the spacetime volume element combined with the generalized temporal gamma matrix as the connecting element.
Section~\ref{sec:RedefFieldTheory} derives a field redefinition which maps the inner product onto its \sMin form and which can be understood in terms of a rescaling factor and local Lorentz boosts.
We then present the redefined Lagrangian, which remains compatible with the underlying diffeomorphism and local-Lorentz gauge symmetries, and transform it into the temporal and spatial decomposition basis, where it takes a generalized canonical Schrödinger form analogous to that in Minkowski spacetime and for a \sMin foliation.
Moreover, we show that this Schrödinger form is directly connected to the redefined Born-rule-like inner product and derive the redefined anti-commutator quantization rule, which also takes a \sMin form.
In Sec.~\ref{sec:Applications} we discuss three possible applications of our formalisms:
First, we highlight that the rescaling replaces the full metric volume element by the lapse function multiplying the rest energy $mc^2$, as expected from the fully covariant Hamiltonian of a classical point particle~\cite{Dodin2010,Poisson2011,Vines2016}.
Second, we show that fixing the local Lorentz frame, in which the generalized temporal gamma matrix is proportional to $\gamma^0$, makes spacetime-magnetic or gravito-magnetic effects, such as those associated with the Lense--Thirring effect~\cite{Lense1918,*Mashhoon1984}, explicit.
Third, we derive the non-relativistic limit of the Lagrangian before and after redefinition.
We find that, before the redefinition, the notion of momentum is less transparent, since the spatially projected derivatives may contain temporal-coordinate derivatives when expressed in non-adapted world coordinates.
In contrast, the redefined Lagrangian assumes a non-relativistic Schrödinger form and a clear distinction between the temporal derivative and spatial momentum, which has direct consequences for first-quantized quantum-mechanical operators.
Finally, in Sec.~\ref{sec:Conclusion} we discuss our results, compare them with previous works, and provide an outlook on possible future extensions and applications.

\section{Quantum field theory for fermions on curved and non-inertial backgrounds}
\label{sec:FieldTheoryBefore}
Our focus lies on fermionic quantum field theories~\cite{Birrell1982,Parker2009,Buchbinder2021} coupled to gauge fields on curved and non-inertial backgrounds.
The metric tensor $g_{\mu\nu}$ describes the spacetime geometry and its components may represent gravity through curved spacetime~\cite{Misner1973} or flat spacetime expressed in non-inertial reference frames or curvilinear coordinates~\cite{Rindler1969,Ni1978,Kajari2009}.
In this work, the metric is treated as a prescribed background and is therefore not necessarily assumed to satisfy the Einstein field equations~\cite{Einstein1915}.
The metric tensor is expressed according to $g_{\mu\nu}={e_\mu}^I{e_\nu}^J\eta_{IJ}$, with the Minkowski metric $\eta_{IJ}$ in the form known from special relativity~\cite{Einstein1905,Rindler1969} in inertial reference frames and Cartesian coordinate systems. 
In the following, we use the mostly-minus signature for the Minkowski metric.
The corresponding objects are the vierbein~\cite{Einstein1930,Birrell1982,Parker2009} field ${E^\mu}_I$ with its dual vierbein ${e_\mu}^I$, which are mutually inverse, \ie{} ${E^\mu}_I{e_\mu}^J=\delta^J_I$ and ${E^\mu}_I{e_\nu}^I=\delta^\mu_\nu$.
Greek indices denote spacetime (world) indices~\cite{Misner1973} and run over four entries. 
While for special foliations introduced later these indices may be interpreted as one temporal and three spatial components, no such distinction is assumed at this stage.
In addition, capital Latin indices label components in local Lorentz frames in which the metric is represented by the Minkowski metric, and run over $0,1,2,3$.
Thus, for the special case of Cartesian inertial coordinates on Minkowski spacetime with a choice ${E^\mu}_I = \delta^\mu_I$, world and local indices coincide. 
The vierbein relates world-coordinate and local-Lorentz bases.
Accordingly, world-index gamma matrices $\gamma^\mu={E^\mu}_I\gamma^I$ are constructed from the constant counterparts $\gamma^I$. 
The vierbein thereby implements the Clifford algebra on curved and non-inertial backgrounds~\cite{Fock1929,Weyl1929,Obukhov2001,Lambiase2021}.
The gamma matrices fulfill~\cite{Peres1962,Birrell1982,Parker2009} the Clifford algebras $\big\lbrace\gamma^{\mu},\gamma^{\nu}\big\rbrace=2g^{\mu\nu}$ and $\left\lbrace\gamma^I,\gamma^J\right\rbrace=2\eta^{IJ}$.

For a spin-$1/2$ fermion coupled to a spin-$1$ gauge field~\cite{Peskin1995,Weinberg1995} on curved and non-inertial backgrounds, the action is defined~\cite{Parker2009} as $S=\int{\mathrm{d}^4x\mathcal{L}}$, with the Lagrangian density
\begin{align}
\label{eq:LagOriginal}
\begin{split}
    \mathcal{L}=\sqrt{-g}\bar{\psi}\left[\frac{\ii\hbar c}{2}\gamma^\mu\lrD_\mu -mc^2\right]\psi
\end{split}
\end{align}
for the fermion sector, including the spinor $\psi$ and the Dirac adjoint spinor $\bar{\psi}=\psi^\dagger\gamma^0$, the fermionic mass $m$, the speed of light $c$, and the determinant $g$ of the metric tensor.
The free gauge-field sector and boundary terms~\cite{Henningson1998,Becker2012,Grosse2025} may be included, but are not necessary for the purpose of our work.
The symmetrized covariant derivative is defined~\cite{Parker2009} by 
\begin{align}
\label{eq:DerSymmBEfore}
\begin{split}
    \bar{\psi}\gamma^\mu\lrD_\mu\psi=\bar{\psi}\gamma^\mu\left(\mathcal{D}_\mu\psi\right)-\left(\mathcal{D}_\mu\bar{\psi}\right)\gamma^\mu\psi,
\end{split}
\end{align}
acting solely on the spinors via
\begin{align}
\label{eq:SymmDer}
\begin{split}
    {\mathcal{D}}_\mu\psi&= \left(\partial_\mu+\frac{\ii}{\hbar} q A_\mu+\Gamma_\mu\right)\psi\\
    {\mathcal{D}}_\mu\bar{\psi}&= \partial_\mu\bar{\psi}-\bar{\psi}\left(\frac{\ii}{\hbar} q A_\mu+\Gamma_\mu\right),
\end{split}
\end{align}
and not on the metric and gamma matrices, which may also depend on world coordinates.
Here, $A_\mu$ is a  spin-$1$ field, possibly carrying additional internal degrees of freedom according to the respective gauge group (\eg{} SU(3) for QCD~\cite{Weinberg1995,Greiner2011}), $q$ its gauge coupling, and $\partial_\mu = \partial/\partial x^\mu$ the partial derivative.
For the spin-$1$ field, gauge invariance requires the one-form $A_\mu$ to transform as a connection such that $\mathcal{D}_\mu\psi$ becomes covariant.
The raised components $A^\mu=g^{\mu\nu}A_\nu$ are derived from the same one-form and are not an independent gauge field~\cite{Hehl2003}.
The spin connection is given by $\Gamma_\mu=\omega_{\mu IJ}\big[\gamma^I,\gamma^J\big]/8$ with $\omega_{\mu IJ}=g_{\nu\lambda}{E^\lambda}_I\big(\partial_\mu {E^\nu}_J+\Gamma^\nu_{\mu\sigma}{E^\sigma}_J\big)$, such that it fulfills the vierbein postulate $\nabla_\mu{E^\nu}_I=\partial_\mu{E^\nu}_I+\Gamma^\nu_{\mu\rho}{E^\rho}_I+{\omega_{\mu I}}^J{E^\nu}_J=0$, with $\Gamma^\nu_{\mu\sigma}$ being the Christoffel symbols~\cite{Weinberg1972,Misner1973}.

In general, the symmetrized minimally-coupled derivative of Eq.~\eqref{eq:SymmDer} provides a manifestly Hermitian form~\cite{Huang2009} of the Lagrangian and fixes the ordering of the gamma matrices. 
In particular, the symmetrized and non-symmetrized expressions are equivalent at the level of the action up to integration by parts, provided that metric and vierbein compatibility hold and the corresponding boundary term vanishes~\cite{Parker2009}.

\subsection{Local-Lorentz-transformation- and diffeomorphism invariances}
Under an LLT, the spinors transform as $\psi^\prime=T\psi$ and $\bar\psi^\prime=\bar\psi T^{-1}$ with $\gamma^0 T^\dagger=T^{-1}\gamma^0$, where the product of world-index gamma matrices and derivatives $\gamma^{\prime\mu}\lrD\,{}^\prime\!\!_\mu=T\gamma^\mu\lrD_\mu T^{-1}$ transforms covariantly.
Here, we defined $T=\exp{\left(\lambda_{IJ}\big(x^{\alpha}\big)\left[\gamma^I,\gamma^J\right]/8\right)}$ as a matrix corresponding to the spinor representation~\cite{Buchbinder2021} of the proper orthochronous Lorentz group~\cite{Tung1985}, with $\lambda_{IJ}\big(x^{\alpha}\big)$ being a real and anti-symmetric matrix, whose dependence on $x^\alpha$ reflects locality.
The vierbeins transform under LLTs as ${E^{\prime\mu}}_{I}={\Lambda_{J}}^I {E^\mu}_J$ and ${e^{\prime}_\mu}^I={\Lambda^{I}}_J{e_\mu}^{J}$, with $\big(\Lambda^{-1}\big)^{I}{}_J={\Lambda_{J}}^I$, where the transforming matrix is $\Lambda(x^\alpha)=\exp{\left(\lambda_{IJ}\big(x^{\alpha}\big)\mathcal{J}^{IJ}/2\right)}$, using the vector generators $\mathcal{J}^{IJ}$ of the Lorentz group~\cite{Tung1985}.
By definition, the Minkowski metric $\eta_{IJ}$ is LLT-invariant at every spacetime point such that ${\Lambda^I}_K{\Lambda^J}_M\eta_{IJ}=\eta_{KM}$.
Since $g_{\mu\nu}={e_\mu}^I{e_\nu}^J\eta_{IJ}$, the metric $g_{\mu\nu}$ is also LLT-invariant, and therefore $\sqrt{-g}$ and the volume element $\mathrm{d}^4x\sqrt{-g}$ are invariant as well.
Hence, the Lagrangian and action are LLT-invariant, such that the choice of the local Lorentz frame at each spacetime point does not modify the physics of the theory.

For diffeomorphisms, \ie{} passive coordinate transformations of four-positions $x^{\mu}\rightarrow \tilde{x}^{\mu}\left(x^{\nu}\right)$,  the vierbein and its dual transform as ${\tilde{E}^{\mu}}{}_I=\left(\partial_\nu\tilde{x}^\mu\right){E^{\nu}}_I$ and ${\tilde{e}_\mu}{}^{J}=\big(\tilde{\partial}_\mu x^\kappa\big){e_\kappa}^{J}$, respectively.
Here, we defined $\partial_\nu\tilde{x}^\mu=\partial \tilde{x}^{\mu}/\partial x^{\nu}$ and $\tilde{\partial}_\mu x^\nu=\partial x^{\nu}/\partial \tilde{x}^{\mu}$.
We find the Jacobian transformation $\mathrm{d}^4\tilde{x}=\big|\det{\partial_\nu\tilde{x}^\mu}\big|\mathrm{d}^4 x$, but since the square root of the determinant $\sqrt{-g}$ transforms as $\sqrt{-\tilde{g}}=\big|\det{\partial_\nu\tilde{x}^\mu}\big|^{-1}\sqrt{-g}$, the full spacetime volume $\mathrm{d}^4x\sqrt{-g}$ is diffeomorphism-invariant.
Since the spinor components transform as scalars under diffeomorphisms, $\mathcal{L}$ transforms as a scalar density, while $\mathrm{d}^4x\,\mathcal{L}$ and hence the action are diffeomorphism-invariant.
Thus the choice of world coordinates does not modify the physical properties of the system.

While the vierbein includes properties of the background geometry through the metric tensor, it is not uniquely and unambiguously determined by the metric.
In addition to the metric degrees of freedom, the vierbein possesses a local Lorentz gauge freedom, corresponding to the LLTs discussed above.
However, many LLTs and diffeomorphisms can be connected by manifold Lorentz transformations~\cite{Woodard1984,Kostelecky2021} $\partial_\nu\tilde{x}^\mu={\Lambda^{\mu}}_\nu$, such that the vierbein transforms as ${\tilde{E}^{\prime\mu}}{}_I={\Lambda^{\mu}}_\nu {\Lambda_I}^J {E^\nu}_J$, connecting a change of world coordinates with a change of the local Lorentz frame.
Manifold Lorentz transformations are the closest analog~\cite{Kostelecky2021} to global Lorentz transformations on Minkowski spacetime, where world and local indices coincide and every Lorentz transformation leads also to a change of world coordinates.

\subsection{Temporal and spatial ADM decomposition}
The metric tensor $g_{\mu\nu}=h_{\mu\nu}+n_{\mu}n_{\nu}$ can be decomposed into a spacelike induced metric $h_{\mu\nu}$ and a timelike normal vector $n_\mu$ being orthogonal to it, with $n_{\mu}n^\mu=1$.
While more general normalizations for $n_\mu$ are possible~\cite{Mars1993}, we focus on spacelike hypersurfaces with future-directed unit timelike normal co-vectors $n_\mu$, as appropriate for the standard spacelike initial-value formulation for massive particles and the positive-definite Dirac inner product.
In the following, we introduce hypersurfaces connected to $h_{\mu\nu}$ and outline their consequences for the theory.

\subsubsection{Hypersurface of simultaneity}
\begin{figure}[!h]
    \centering
    \includegraphics[width=\columnwidth]{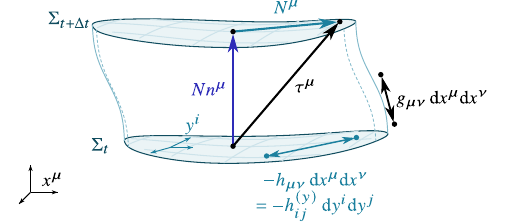}
    \caption{
    A point on the hypersurface $\Sigma_t$ evolves to its location on $\Sigma_{t+\Delta t}$ through the evolution vector $\tau^\mu$, which can be separated into a temporal part $N n^\mu$ and a shift $N^\mu$.
    While the metric $g_{\mu\nu}$ measures general spacetime intervals through $g_{\mu\nu}\mathrm{d}x^\mu \mathrm{d}x^\nu$, the induced metric measures spatial distances on the hypersurface through $-h_{\mu\nu}\mathrm{d}x^\mu \mathrm{d}x^\nu$, which can be projected onto spatial coordinates $y^i$ only valid on the hypersurface via $-h^{(y)}_{ij}\mathrm{d}y^i \mathrm{d}y^j$.
    }
    \label{fig:GeneralHypersurface}
\end{figure}

We consider spacetimes admitting a smooth foliation by spacelike hypersurfaces $\Sigma_t$ connected to $h_{\mu\nu}$, with $n_\mu$ denoting their future-directed unit normal co-vector.
Thus, $\Sigma_t$ is part of a family of hypersurfaces, each represented by a fixed value of the foliation-time function $t(x^\mu)=\mathrm{const.}$. 
In general, the normal vector and hypersurface do not need to be directly connected to a temporal direction and simultaneity.
The foliation-time function $t(x^\mu)$ can be an arbitrary function of world coordinates and is in general not equal to $x^{\bar{0}}/c$.
Its gradient is assumed to be timelike everywhere and non-vanishing in the region considered.
This form of metric decomposition is called Arnowitt–Deser–Misner (ADM) decomposition~\cite{Arnowitt1960,Arnowitt2008}.
All points on a given hypersurface $\Sigma_t$ are simultaneous with respect to the chosen foliation and can be labeled by (abstract) spatial coordinates.
Hence, the tensor $h_{\mu\nu}=g_{\mu\nu}-n_\mu n_\nu$ projects onto directions tangent to $\Sigma_t$ and its pullback to the hypersurface defines the induced spatial metric.
Consequently, the normal co-vector on the hypersurface $n_\mu = N \partial_\mu c t(x)$ determines the normal temporal direction associated with the foliation.
As such, the normal vector depends on the metric and coordinate system but includes still additional degrees of freedom, as explained later.
The normalization $N$ is the lapse function~\cite{Arnowitt1960}, defining how the coordinate time $t(x)$ flows in given coordinates and reference frames. 
From the condition $n_\mu n^\mu =1$ we find that the lapse function is $N = 1/\sqrt{ c^2 g^{\mu \nu} \partial_\mu t \partial_\nu t}$.
Thus, the simultaneity of two events is in general not necessarily defined through the zeroth coordinate in a given reference frame for arbitrary $n_\mu$, but rather through the condition $t(x^\mu)=\text{const.}$ imposed by the hypersurface.
Moving in time, points on the hypersurface $\Sigma_t$ are connected to later and shifted points on $\Sigma_{t+\Delta t}$ by the evolution vector $\tau^\mu \left( x \right) = \partial x^\mu / \partial \left( c t \right) = N n^\mu + N^\mu$, which can be separated into a temporal direction through $n^\mu$ and shifts on the hypersurface according to the shift vector $N^\mu$, with $n_\mu N^\mu=0$, see Fig.~\ref{fig:GeneralHypersurface}.
On the hypersurface, we define the projection $h^{(y)}_{ij}=h_{\mu\nu}\big(\partial^{(y)}_i\!x^\mu\big)\big(\partial^{(y)}_j\!x^\nu\big)$, with $\partial^{(y)}_i\!x^\mu=\partial x^\mu/\partial y^i$, of the spacelike induced metric $h_{\mu\nu}$ onto a spatial coordinate system\footnote{
Introducing hypersurface quantities does not necessarily require expressing these quantities in explicit $y$ coordinates.
One can always project in world coordinates $x^\mu$ onto the hypersurface basis, such that, \eg $h^{(y)}_{ij} = h^{(y)}_{ij} \left( x^\mu \right)$ is obtained.
} labeled by $y^i$, indicated by lowercase Latin indices, like those used for the last three indices of world coordinates $\mu$.
The spatially projected vierbein defines a dreibein on each hypersurface, which is one of the geometric ingredients entering the Ashtekar formulation~\cite{Ashtekar1986}.
As a consequence, an observer measures spatial distances according to $-h^{(y)}_{ij}\mathrm{d}y^i\mathrm{d}y^j$ in the $y^i$-coordinate projection, while these spatial distances depend on the hypersurface and in general on all spacetime coordinates.
Note that $h_{ij}$ is equal to $g_{ij}$ if and only if $n_\mu$ points solely in zeroth direction.
In Fig.~\ref{fig:GeneralHypersurface} we illustrate the role of evolution vector, normal vector, and shift vector, as well as the connection of world coordinates and projected spatial coordinates through the induced metric.

\subsubsection{Examples for hypersurfaces of simultaneity}
To illustrate the notion of spacetime foliations, we assume an arbitrary but fixed metric and coordinate system, and embed different hypersurfaces $\Sigma_t$ into these curved or non-inertial backgrounds by choosing a particular foliation-time function $t\left( x^\mu \right)$.
For a given hypersurface, there are infinitely many possible observers which are defined by a timelike four-velocity $U^\mu$ , describing the observer's motion through spacetime.
Two prime examples~\cite{Gourgoulhon2012,Alcubierre2025,Teukolsky2026} are given by the coordinate observer $U^\mu \parallel \tau^\mu$, provided that $\tau^\mu$ is timelike, and the Eulerian observer $U^\mu = n^\mu$.
Hence, the way in which the hypersurface is embedded into curved or non-inertial spacetime determines the Eulerian observer in these fixed coordinates. 
With respect to a chosen foliation, the Eulerian observer measures the normal Dirac-current density $U_\mu j^\mu =n_\mu j^\mu = j_t$ (and orthogonally to it a spatial current) with the four-current $j^\mu=\bar{\psi}\gamma^\mu \psi$, while in general a measurement of a four-current $V_\mu j^\mu=V_t j_t / N+V^{(y)}_i j^{(y)i}$, with $V_t=V_\mu\tau^\mu$, $V^{(y)}_i=V_\mu\partial^{(y)}_i\!x^\mu$, and $j^{(y)i}=j^\nu\partial_\nu y^i$, from any other observer $V^\mu$ is a mixture of temporal density and spatial current. 

\paragraph{Constant-foliation-time foliation}
The prime example is the foliation-time choice $t(x^\mu)=x^{\bar{0}}/c$.
Here, we write for the zeroth component of world indices $\mu$ now ${\bar{0}}$, to distinguish it from the zeroth coordinate $0$ of the local indices $I$.
This example treats $x^{\bar{0}}$ indeed as coordinate time and represents constant-foliation-time foliation characterized by $n_\mu=N \delta^{\bar{0}}_\mu$ and $n^\mu= N g^{\mu {\bar{0}}}$, where the lapse function takes the simple and established~\cite{Misner1973} form $N=\big(g^{{\bar{0}}{\bar{0}}}\big)^{-1/2}$. 
With respect to the fixed coordinates, the evolution vector is $\tau^{\mu}=\left( 1, \partial x^i / \partial\left( c t \right) \right)$ and the shift vector reduces to $N^{\mu}=\left(0, \partial x^i / \partial\left( c t \right) - N^2 g^{i \bar{0}} \right)$.
It takes into account the difference between the Eulerian and the coordinate observer, which differ either when moving in time within the fixed coordinates implies a spatial drift induced by $g^{i \bar{0}}$ or when the coordinate lines $x^i$ are moving over time induced by the choice of foliation $\partial x^i/\partial t$.
For example, in a foliation with $\partial x^i/\partial t = 0$, the coordinate observer takes the simple form $\tau^\mu = \delta^\mu_{\bar0}$, and in a coordinate system where $g^{i \bar{0}}=0$, the Eulerian observer has the simple form $n^\mu = \big(g^{\bar{0}\bar{0}}\big)^{1/2}\delta^\mu_{\bar0}$.
If both cases apply, Eulerian and coordinate observer coincide.
To complete the foliation-adapted coordinate system and fix the threading between neighboring hypersurfaces, a choice must be made for $y^i(x^\mu)$ or $x^i(t,y^i)$. 
A simple choice is $x^i(t,y^i)=y^i$, for which the spatial labels are identified with the corresponding world coordinates and remain fixed along the foliation-time lines.

\paragraph{Co-moving hypersurface in Minkowski spacetime}
In the special case of Cartesian inertial coordinates on Minkowski spacetime, it is often convenient to define a spacetime foliation $t(x) = v_I x^I/c^2$ with respect to a constant four velocity $v_I$ with $v_I v^I = c^2$. 
As a consequence, the hypersurface normal co-vector is defined by $n_I = v_I/c$ with a trivial lapse function $N=1$, generalizing the case $n_I=\delta^{0}_I$ to hypersurfaces that are tilted by $v_I$.
This Eulerian observer $U^I \parallel v^I$ can be interpreted as being attached to an inertial frame moving with constant four-velocity $v^I$ relative to the laboratory frame defined by the $n^I=\delta^I_{0}$ direction. 
Besides fixing a foliation-time function $t(x)$, we have in this case the freedom to determine the dependence of $y^i(x)$ on the fixed coordinates, which defines then the remaining quantities of the basis, like the evolution vector $\tau^I$ and the shift vector $N^\mu$.

\paragraph{Hypersurfaces attached to observers}
For a given coordinate system and metric in curved or non-inertial backgrounds, it is often convenient to choose a spacetime foliation such that the Eulerian observer, defined by the corresponding hypersurface, aligns with another observer, \eg{} the observer co-moving with a particle of non-constant four-velocity $V^\mu$ or a freely falling observer.
In many cases, an explicit foliation-time function may not be readily available. 
One may instead attempt to align the normal vector with a given observer or a class of equivalent observers, $n^\mu\parallel V^\mu$, and construct the lapse, shift, and hypersurface basis consistently. 
However, such an identification is possible only if the equivalent-observer class is orthogonal to the hypersurface.
Hence, not every observer can define the Eulerian observers of a foliation.

\subsection{Lagrangian, inner product, and quantization in ADM basis}
In the following, we express the structures of our fermionic theory, such as the Lagrangian, inner product, and quantization rule, in the ADM decomposition to identify the relevant redefinition afterwards.
To this end, we consider the vector basis $\{\tau^\mu,\partial^{(y)}_i\!x^\mu\}$ and the dual co-vector basis $\{ \partial_\mu c t, \partial_\mu y^i \}$.
Hence, they fulfill the orthogonality relations $\tau^\mu\partial_\mu c t(x)=1$, $\big(\partial^{(y)}_i\!x^\mu\big)\partial_\mu y^j=\delta^j_i$, $\big(\partial^{(y)}_i\!x^\mu\big)\partial_\mu c t(x)=0$, and $\tau^\mu \partial_\mu y^i =0$, yielding the completeness relation $\delta^\mu_\nu = \tau^\mu \partial_\nu c t+\big(\partial^{(y)}_i\!x^\mu\big)\partial_\nu y^i$ that allows to express every contravariant and covariant quantity with respect to these vector and co-vector bases.
For example, defining the contravariant projected induced metric as $h^{(y)ij}=h^{\sigma\rho}\big(\partial_\sigma y^i \big)\big(\partial_\rho y^j\big)$, the relation $h^{\mu \nu} = {h^{(y)ij}} \big( \partial^{(y)}_i x^\mu \big) \big( \partial^{(y)}_j x^\nu \big)$ follows because $h^{\sigma \rho}\partial_\sigma c t=0$.
Similarly, with $h^{(y)}_{ij}=h_{\sigma\rho}\big(\partial^{(y)}_i\!x^\sigma\big)\big(\partial^{(y)}_j\!x^\rho\big)$ we find $h_{\mu \nu} =h^{(y)}_{ij} \big[N^{(y)i} n_\mu / N + \big(\partial_\mu y^i\big) \big] \big[N^{(y)j} n_\nu / N  + \big(\partial_\nu y^j \big) \big]$, where the projected shift vector $N^{(y)i} = N^\mu \partial_\mu y^i$ appears because $\tau^\mu \partial_\mu y^i =0$ holds only along the time-evolution vector $\tau^\mu$ and not along $n^\mu$. 
The projected induced metric obeys the standard relation ${h^{(y)ij}}h^{(y)}_{jk} = \delta^i_k $. 

Using these bases, we will show that the ADM decomposed Lagrangian naturally determines the inner product and the anti-commutator quantization rule.
Especially, all three objects contain the temporal gamma matrix $\gamma_t=n_\mu\gamma^\mu/N=\gamma^\mu\partial_\mu c t(x)$.

\subsubsection{Lagrangian before redefinition in ADM basis}
Using $\delta^\mu_\nu = \tau^\mu n_\nu/N + \big(\partial^{(y)}_i\!x^\mu\big)\big(\partial_\nu y^i\big)$, we recast the original Lagrangian from Eq.~\eqref{eq:LagOriginal} in the ADM basis as
\begin{align}
\label{eq:LagADM}
\begin{split}
    \mathcal{L}=\sqrt{-g}\bar{\psi}\left[\frac{\ii\hbar c}{2}\left(\gamma_t\lrD{}^{(\tau)}_t+\gamma^{(y)i}\lrD{}^{(y)}_i\right)-mc^2\right]\psi,
\end{split}
\end{align}
with temporal gamma matrix $\gamma_t=n_\mu\gamma^\mu/N$, temporal derivative $\lrD{}^{(\tau)}_t=\tau^\nu\lrD_\nu$, spatial gamma matrix $\gamma^{(y)i}=\gamma^\mu\partial_\mu y^i$, and spatial derivative $\lrD{}^{(y)}_i=\big(\partial^{(y)}_i\!x^\nu\big)\lrD_\nu$.
The gamma matrices in the new basis fulfill $\left\lbrace\gamma_t,\gamma_t\right\rbrace=2/N^2$, $\left\lbrace\gamma_t,\gamma^{(y)i}\right\rbrace=-2N^{(y)i}/N^2$, and $\left\lbrace\gamma^{(y)i},\gamma^{(y)j}\right\rbrace=2\left(h^{(y)ij}+N^{(y)i} N^{(y)j}/N^2\right)$, which just corresponds to $g^{\mu \nu}$ projected onto the vector and co-vector bases.
Hence, we define new ADM indices $\Omega,\Xi\in \lbrace t,i\rbrace$ with capital Greek letters, yielding the compact form $\left\lbrace\gamma^{(y)\Omega},\gamma^{(y)\Xi}\right\rbrace=2g^{\left(y\right)\Omega\Xi}$, which still fulfills a Clifford algebra in this representation with respect to the projected ADM metric $ g^{\left(y\right)\Omega\Xi} = g^{\mu\nu} \big(\partial_\mu y^\Omega\big) \big(\partial_\nu y^\Xi\big)$, where $\gamma^{(y)t}=\gamma_t$ and $y^t = ct$.
Thus, these gamma matrices are constructed from the hypersurface-projected vierbein and thereby inherit the relation of its spatial part to the dreibein entering the Ashtekar formulation~\cite{Ashtekar1986}.
In complete analogy to the original basis, the Lagrangian in this new basis remains LLT- and diffeomorphism-invariant.
Although $\gamma_t$ represents a generalized temporal matrix, the naive Schrödinger form $\ii\hbar\psi^\dagger\lrpartial_t\psi/2$ is lost in non-inertial backgrounds due to $\gamma_t$ and $\sqrt{-g}$.
Hence, the Lagrangian and equations of motion do not directly display the standard Schrödinger form with a unit temporal kinetic kernel, obscuring the interpretation of the background contributions as additional effective interactions.

\subsubsection{Non-inertial inner product before redefinition}
Any physically valid fermionic inner product~\cite{Leclerc2006} requires an LLT- and diffeomorphism-invariant definition and must provide a consistent notion of the norm of the states.
Together with $\nabla_\mu\psi={\mathcal{D}}_\mu\psi-\ii q A_\mu\psi/\hbar$, $\nabla_\mu\gamma^\mu=0$, and $\nabla_\mu\sqrt{-g}=0$, we obtain from varying~\cite{Parker2009} the Lagrangian in Eq.~\eqref{eq:LagOriginal} through $\partial\mathcal{L}/\partial\bar\psi-\nabla_\mu\left(\partial\mathcal{L}/\partial\left(\nabla_\mu\bar\psi\right)\right)=0$ the non-inertial Dirac equation~\cite{Parker1980,Obukhov2001,Obukhov2011,Lambiase2021,Alcubierre2025} with respect to $\psi$ as $\ii\hbar c\gamma^\mu\mathcal{D}_\mu\psi=mc^2\psi$, and analogously the adjoint Dirac equation with respect to $\bar\psi$. 
Upon summation, they lead to $\sqrt{-g}\nabla_\mu\left(\bar{\psi}\gamma^\mu\psi\right)=\partial_\mu\left(\sqrt{-g}\bar{\psi}\gamma^\mu\psi\right)=0$ on shell.
Integrating over a spacetime region $\mathcal{V}$ bounded by two hypersurfaces $\Sigma_{t_1}$ and $\Sigma_{t_2}$ and a lateral boundary $\mathcal{B}$, Gauss's theorem yields
\begin{align}
\begin{split}
    0=& \int \limits_\mathcal{V} \mathrm{d}^4 x \partial_\mu\left(\sqrt{-g}\bar{\psi}\gamma^\mu\psi\right) \\
    =& \int_{\Sigma_{t_2}}\!\!{\mathrm{d}\Sigma_\mu \bar{\psi} \gamma^\mu\psi}
    -\int_{\Sigma_{t_1}}\!\!{\mathrm{d}\Sigma_\mu \bar{\psi}\gamma^\mu\psi}
    + \int_{\mathcal{B}}\!\!\mathrm{d}\Sigma_\mu\,\bar\psi\gamma^\mu\psi.
\end{split}
\end{align}
On each spacelike hypersurface, we use the future-directed hypersurface element $\mathrm{d}\Sigma_\mu=\mathrm{d}\Sigma n_\mu$.
The explicit minus sign in front of $\Sigma_{t_1}$ accounts for its opposite orientation as part of the boundary $\partial\mathcal{V}$. 
On the lateral boundary $\mathcal{B}$, $\mathrm{d}\Sigma_\mu$ denotes the outward-directed hypersurface element.
Assuming that the flux through $\mathcal{B}$ vanishes~\cite{Parker2009}, the integrals over $\Sigma_{t_1}$ and $\Sigma_{t_2}$ are equal.
To display the factor $\sqrt{-g}\gamma_t$ found from the Lagrangian explicitly, we introduce the coordinate hypersurface measure $\mathrm{d}\sigma$ through $\int_{\Sigma_t} \mathrm{d}\sigma \, G = \int \mathrm{d}^4 x \, \delta( c t(x) - ct ) G(x)$, such that the identification $\mathrm{d}\Sigma = \mathrm{d}\sigma \sqrt{-g} / N$ of the intrinsic hypersurface element follows.
Note that $\mathrm{d}\sigma$ contains the foliation-time function $t(x)$ as a pure function of world coordinates $x^\mu$, while $t$ without an argument corresponds to the fixed parameter labeling the respective hypersurface $\Sigma_t$.
In explicit, \ie{} hypersurface-adapted, $y$ coordinates, we find $\mathrm{d}\Sigma = \mathrm{d}^3y \sqrt{-h}$, where $\sqrt{-h}$ is the square root of the negative determinant of $h^{(y)}_{ij}$.
In these coordinates, the delta distribution $\delta( c t(x) - ct )$ is, upon integration, exactly responsible for $\mathrm{d}^4x$ reducing to $\mathrm{d}^3y$ and all functions become dependent on the fixed hypersurface time $t$.
Hence, the fermionic current $j^\mu$ integrated over $\Sigma_t$ naturally defines~\cite{Leclerc2006} the fermionic inner product
\begin{align}
\begin{split} \label{eq:innerprod} 
    \left(\psi_1,\psi_2\right)&=\int_{\Sigma_t}\!\!{\mathrm{d}\sigma\sqrt{-g}\bar{\psi}_1\gamma_t\psi_2}=\int_{\Sigma_t}\!\!{\mathrm{d}^3y\sqrt{-g^{(y)}}\bar{\psi}_1\gamma_t\psi_2}\\
    &=\int_{\Sigma_t}\!\!{\mathrm{d}^3y\sqrt{-h}\bar{\psi}_1n_\mu\gamma^\mu\psi_2}
\end{split}
\end{align} 
between two spinors $\psi_1$ and $\psi_2$ in curved or non-inertial backgrounds, for which we used $\sqrt{-g^{(y)}}=N\sqrt{-h}$ valid on the spacetime foliation within the hypersurface coordinates, where $g^{(y)}=\det{g^{\left(y\right)}_{\Omega\Xi}}$ is the determinant of the projected metric tensor, and we use $\psi=\psi(x^\mu(y^\Xi))$ in integrals over explicit $y$ coordinates\footnote{Whenever we use quantities such as $\sqrt{-g}$, $\sqrt{-g^{(y)}}$, $\sqrt{-h}$, or $\psi$ in y-coordinate integrals, we implicitly assume them as evaluated at $f(x^\mu)=f(x^\mu(y^\Xi))$, without introducing new symbols.}.
This definition is directly connected to the future-directed temporal projection $n_\mu j^\mu$ and yields a constant normalization of physical states.
Thus, the term involving the temporal derivative in the Lagrangian and the inner product are directly connected.
This connection implies, however, that the non-inertial inner product does not obey the \sMin Born-rule form $\bar{\psi}\gamma^0\psi$ or $\psi^\dagger\psi$.
As such, the inner product possesses the Hermitian kernel $\sqrt{-h}n_\mu\gamma^0\gamma^\mu$.
Moreover, the inner product itself does not need to be restricted to the special choice $\Sigma_t$, but can in principle be defined on any suitable spacelike Cauchy hypersurface.
However, for the initial-value problems considered here, $\Sigma_t$ is a natural choice and leads to a positive definite inner product.
Since $n_\mu\gamma^\mu$ transforms covariantly under LLTs and as a world-coordinate scalar under diffeomorphisms, the complete hypersurface integral, including its invariant measure, is LLT- and diffeomorphism-invariant.
Moreover, since the fermionic current is a conserved Noether current~\cite{Fulling1989,Fatibene2003}, the inner product does not depend on the chosen hypersurface, provided that the relevant boundary flux vanishes. 
The norm of states does also not depend on the choice of coordinates due to the covariant hypersurface integral.

\subsubsection{Non-inertial anti-commutator before redefinition}
The Lagrangian from Eq.~\eqref{eq:LagADM} and the inner product or bilinear form from Eq.~\eqref{eq:innerprod} are connected to the quantization rule of the fermion fields.
The connection to the Lagrangian can be observed from the canonical momentum as $\pi_\psi=\partial\mathcal{L}/\partial\left( \partial_t \psi\right)=\partial\mathcal{L}/\partial\left(c\tau^\nu\partial_\nu\psi\right)$, which is the derivative of the Lagrangian density with respect to the field velocity defined by the chosen foliation.
We obtain the canonical momentum $\pi_\psi=\ii\hbar\sqrt{-g}\bar{\psi}\gamma_t/2$ and $\pi^{(y)}_{\psi}=\ii\hbar\sqrt{-h}\bar{\psi}n_\mu\gamma^\mu/2$ in world coordinates $x^\mu$ and in hypersurface-adapted coordinates $y^\Omega$, respectively, which take the form of the kernel appearing in the inner product.
This derivation can be performed analogously for $\pi_{\bar{\psi}}$ (or $\pi_{\psi^\dagger}$) in both coordinates.
With that result, the symmetrized bilinear form $\left(\psi,\psi\right)=\int{\mathrm{d}^3y\big(\pi^{(y)}_\psi\psi-\bar{\psi}\pi^{(y)}_{\bar{\psi}}\big)/(\ii\hbar)}$ connects the inner product directly to conjugate momenta. 
Hence, we postulate the canonical Poisson bracket~\cite{Nelson1978} $\left.\left\lbrace\psi_a(x),\pi_{\psi,b}(x^\prime)\right\rbrace_\text{P} \right|_{\Sigma_t} =\delta_{ab} \delta_{\sigma}\big( x, x^\prime \big)$, where small Latin indices indicate now spinor components and not spacetime indices and the delta distribution $\delta_{\sigma}(x,x^\prime)$ satisfies $\int_{\Sigma_t} \mathrm{d}\sigma(x^\prime) \delta_{\sigma} \big( x, x^\prime \big) F\big( x^\prime \big) = F (x)$ if $x^\mu$ lies on $\Sigma_t$, with respect to the measure $\mathrm{d}\sigma(x^\prime) = \mathrm{d}^4x^\prime \delta ( ct (x^\prime) - ct )$ as before.
Based on how we define the conjugate momentum derived from the Lagrangian in world coordinates, the Poisson bracket has to reproduce a function $\int \mathrm{d}\sigma(x^\prime) \left\{ \psi_a (x), \pi_{\psi,b} \big(x^\prime \big) \right\}_\text{P} F_b(x^\prime ) = F_a (x)$.
An analogous Poisson bracket is valid for $\bar{\psi}$ and $\pi_{\bar{\psi}}$, while other combinations vanish.
From the second-class constraints~\cite{Leclerc2007} $\pi_\psi-\ii\hbar\sqrt{-g}\bar{\psi}\gamma_t/2\approx0$ and $\pi_{\bar{\psi}}+\ii\hbar\sqrt{-g}\gamma_t\psi/2\approx0$, we obtain in $x$ coordinates the Dirac bracket $\left.\left\lbrace\psi_a(x),\bar\psi_b(x')\right\rbrace_\mathrm D\right|_{\Sigma_t}= N^2\gamma_{t,ab}\delta_{\sigma}(x,x')/(\ii \hbar \sqrt{-g})$,
which can be quantized as $\ii\hbar\left\lbrace\psi_a,\bar{\psi}_b\right\rbrace_\text{D}\rightarrow\left\lbrace\psi_a,\bar{\psi}_b\right\rbrace$.

As a consequence, we find the quantized anti-commutator relations
\begin{align}
\begin{split}
    \label{eq:QuantADM}
    \left.\left\lbrace\psi_a\big(x\big),\bar{\psi}_b\big(x^\prime\big)\right\rbrace\right|_{\Sigma_t}=& \frac{N}{\sqrt{-g}} n_\mu\gamma^\mu_{ab}\delta_{\sigma}\left(x,x^\prime\right) \\
    =&\frac{N^2}{\sqrt{-g}} \gamma_{t,ab}\delta_{\sigma}\left(x,x^\prime\right)
\end{split}
\end{align}
on curved or non-inertial backgrounds at equal-temporal spacetime points in $x$ world coordinates\footnote{In explicit $y$ coordinates, the anti-commutator translates to $\left\lbrace\psi_{a}\left(t,\vect{y}\right),\bar{\psi}_{b}\left(t,\vect{y}^{\prime }\right)\right\rbrace=N\gamma_{t,ab}\delta^{(3)}\left(\vect{y}-\vect{y}^{\prime }\right)/\sqrt{-h}$, highlighting the equal-time nature of the quantization rule.} for spinor components $a$ and $b$ and for two points $x$ and $x^\prime$ on the same hypersurface $\Sigma_t$ of the chosen foliation.
In the ADM decomposition, the anti-commutator is governed by the temporal gamma matrix $\gamma_t$, corresponding to a generalized notion of an equal-time quantization rule.
So far, the Dirac bracket (and thus the anti-commutator) for fermions depends explicitly on non-inertial quantities.
Consequently, the fermionic quantization rule retains an explicit dependence on the spacetime background and the chosen foliation, which may lead to ambiguities~\cite{Morales1994,Thiemann1998} when quantizing both.

\section{Recovering \sMin structures through spinor field redefinition}
\label{sec:RedefFieldTheory}
In many experiments, observables constructed from the fermion field are measured~\cite{Charpak1991,Moser2009,Malberti2026} directly.
Consequently, as outlined in the introduction Sec.~\ref{sec:intro}, it is often desirable to recover the corresponding properties in their \sMin form.
Such a form is particularly convenient for computing expectation values, transition matrix elements, scattering amplitudes, particle numbers, or related quantities.
As derived in the previous section in Eqs.~\eqref{eq:LagADM} and \eqref{eq:innerprod}, the deviation from this standard structure arises from the factor $\sqrt{-g}\gamma_t$.
Hence, the time evolution does not assume Schrödinger form $\ii\hbar\psi^\dagger\lrpartial_t\psi/2$, and the identification of excitations from the inner product or bilinear form depends on world coordinates and the inertial frame.
Therefore, $\sqrt{-g}\bar{\psi}\gamma_t\psi$ may not in general be interpreted as a particle density of the form $\bar{\psi} \gamma^0 \psi$, but rather as the densitized normal projection $\sqrt{-g}\,n_\mu j^\mu/N$ of the Dirac current, determined by the background and the chosen foliation.

To solve these issues, we construct a field redefinition~\cite{Kamefuchi1961,Peskin1995,Weinberg1995,Burgess2020} $\psi=K\chi$ and $\psi^\dagger=\chi^\dagger K$, with Hermitian $K$ defined below, such that $\sqrt{-g}\bar{\psi}\gamma_t\psi=\bar{\chi}\gamma^0\chi$.
As a consequence, we find $\sqrt{-g}\bar{\psi}\gamma_t\lrpartial_t\psi=\bar{\chi}\gamma^0\lrpartial_t\chi$ with $\bar{\chi}=\chi^\dagger\gamma^0$.
This transformation maps the inner product and canonical normalization to their \sMin forms while transferring their explicit background dependence to the transformed Lagrangian and field operators, thus allowing for computations to be made in their well-known \sMin form.
Thus, for an invertible Hermitian $K$, the defining equation is $\sqrt{-g} \gamma^0 \gamma_t = K^{-2}$, such that determining $K$ requires the inversion and matrix square root of this expression.
Since $\det{K^{-2}} = (-g)^2/N^4\neq0$, the matrix $K^2$ exists and is uniquely determined.
For a future-directed hypersurface normal co-vector and a time-oriented vierbein with $E^{(n)}_0>0$, $K^{-2}$ and hence $K^2$ are Hermitian and positive definite. 
Thus, $K^2$ has a unique positive-definite Hermitian square root $K$, while unrestricted Hermitian square roots can possess additional branches.
Restricting the Hermitian roots to local covariant Clifford structures constructed solely from the matrix $K^{-2}$, we obtain there are two physically distinct solutions\footnote{There are two other solutions $-K$ with a global minus sign. Since $K$ appears always squared in the inner product and the Lagrangian, they have no distinct physical effect compared to the global positive solutions.} for $K$ denoted by $\pm$, which resolve to
\begin{align}
\begin{split}
    K=&\left(-g\right)^{-1/4}\sqrt{N}C^{\pm}_I\gamma^0\gamma^{I},
\end{split}
\end{align}
with real $C^{\pm}_0 = ( E_0^{(n)} \pm1 )^{1/2} / \sqrt{2}$, $C^{\pm}_{\hat{i}}= - (E_0^{(n)} \mp 1)^{1/2} E_{\hat{i}}^{(n)} / \big( E^{(n)} \sqrt{2} \big)$.
Here, we define the projection $E^{(n)}_I=n_\mu {E^\mu}_I$ of the vierbein in $n_\mu$-direction with $E^{(n)}=\sqrt{-\eta^{\hat{i}\hat{j}} E^{(n)}_{\hat{i}} E^{(n)}_{\hat{j}}}\geq0$.
Note that the indices $\hat{i}=1,2,3$ are the last three local indices $I$, which are in general different from the last three world indices $i$ of $\mu$.
Moreover, we find the relation $C^{\pm}_I C^{\pm}_J \eta^{IJ} =\pm1$.
By inserting this transformation, we directly find the desired Schrödinger form of the Lagrangian and the Born-rule form of inner product, as shown in the following.

The redefinition $K$ can be decomposed into a rescaling by $\left(-g\right)^{-1/4}\sqrt{N}$ and a matrix-valued transformation $C^{\pm}_I\gamma^0\gamma^{I}$.
The effect of the rescaling on the Lagrangian amounts, due to the symmetrized derivative, to replacing the factor $\sqrt{-g}$ by the lapse function $N$.
The same holds for the inner product.
In addition, this transformation maps the background-dependent matrix kernel in the anti-commutation relation to $\gamma^0$.
This way, the rest energy $N mc^2$ akin to a classical point particle is restored through the rescaling part of the redefinition, as discussed later.
While the original spinor $\psi$ is invariant under diffeomorphisms, we find that $\chi$ transforms as $\tilde{\chi}=\sqrt{\big|\det{\partial_\nu\tilde{x}^\mu}\big|^{-1}}\chi$ in $x$ coordinates since $\tilde{N}=N$.

Apart from the rescaling, the remainder of the redefinition can be understood as an LLT boost
\begin{align}
\begin{split}
    {T_{\gamma^0}}^{-1}\equiv C^{+}_I\gamma^0\gamma^{I} = \exp{ \left\{ \frac{\vartheta_{\hat{i}}}{2} \gamma^0 \gamma^{\hat{i}} \right\} }, 
\end{split}
\end{align}
for the plus solution, with rapidity divided by its norm $\vartheta_{\hat{i}} / \text{Arcosh} \, E^{(n)}_0 = - E^{(n)}_{\hat{i}} / E^{(n)}$.
The minus solution can be expressed as $E^{(n)}_{\hat{i}}/E^{(n)} \gamma^{0} \gamma^{\hat{i}} {T_{\gamma^0}}^{-1}$ for $E^{(n)}>0$.
The branch associated with the minus solution reverses the sign of the transformed rest-energy term and yields the opposite sign relative to the positive lapse. 
Restoring the standard form would require $N\rightarrow -N$, which reverses the orientation of $n_\mu=N\partial_\mu(ct)$ for fixed $t(x)$.  
Since we fix $N>0$ and a future-directed hypersurface normal, we restrict the following construction to the positive branch.

\begin{figure}[!h]
    \centering
    \includegraphics{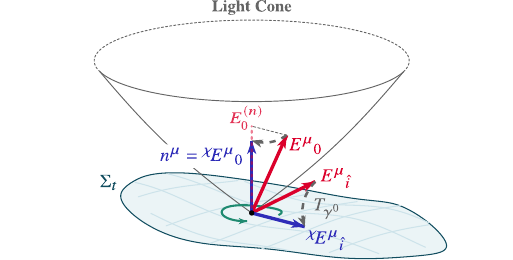}
    \caption{The tangent space at a point of the hypersurface $\Sigma_t$ (tinted, see Fig.~\ref{fig:GeneralHypersurface}) is shown in the rest frame of the Eulerian observer $U^\mu=n^\mu$, such that the future light cone is symmetric about the normal $n^\mu$.
    Before the redefinition (red), a generic vierbein has a timelike leg ${E^\mu}_0$ tilted against $n^\mu$, with projection $E^{(n)}_0=n_\mu{E^\mu}_0$ (red dashed line), and spatial legs ${E^\mu}_{\hat{i}}$ that are not tangent to $\Sigma_t$, since $n_\mu{E^\mu}_{\hat{i}}=E^{(n)}_{\hat{i}}\neq0$.
    The rotation-free LLT boost $T_{\gamma^0}$ (dashed arrows), with rapidity $\text{Arcosh}\,E^{(n)}_0$ directed along $-E^{(n)}_{\hat{i}}/E^{(n)}$, maps the frame onto the Schwinger time gauge (blue), ${}^\chi\hspace{-0.22em}{E^\mu}_0=n^\mu$ and $n_\mu{}^\chi\hspace{-0.22em}{E^\mu}_{\hat{i}}=0$, in which $N\gamma_t$ is mapped to $\gamma^0$.
    This choice is unique up to spatial rotations (green circular arrow) within $\Sigma_t$, which leave $\gamma^0$ invariant.
    }
    \label{fig:FrameFixing}
\end{figure}

With this special LLT, fixing a local Lorentz frame, we obtain $N T_{\gamma^0}\gamma_t {T_{\gamma^0}}^{-1}=\gamma^0$ with $T_{\gamma^0} \gamma^J {T_{\gamma^0}}^{-1} =  {}^{\gamma^0}\hspace{-0.3em}{\Lambda_I}^J \gamma^I $ and $ {}^{\gamma^0}\hspace{-0.3em}{\Lambda_I}^J E^{(n)}_J = \delta^0_I$, where the explicit boost components are given by ${}^{\gamma^0}\hspace{-0.3em}{\Lambda_0}^0 = E^{(n)}_0$, ${}^{\gamma^0}\hspace{-0.3em}{\Lambda_{\hat{i}}}^0=- E^{(n)}_{\hat{i}}$, ${}^{\gamma^0}\hspace{-0.3em}{\Lambda_0}^{\hat{i}} = \eta^{\hat{i} \hat{j}} E^{(n)}_{\hat{j}}$, and ${}^{\gamma^0}\hspace{-0.3em}{\Lambda_{\hat{i}}}^{\hat{j}} = \delta^{\hat{j}}_{\hat{i}}- E^{(n)}_{\hat{i}} E^{(n)}_{\hat{k}} \eta^{ \hat{k} \hat{j}} / \left(E^{(n)}_0+1\right)$.
This choice for the local-Lorentz-frame fixing, representing the Schwinger-time gauge~\cite{Schwinger1963} locally in arbitrary spacetimes and for arbitrary spacelike hypersurfaces, always implies that ${}^{\gamma^0}\hspace{-0.3em}{\Lambda_0}^J {E^\mu}_J=n^\mu$.
It therefore imposes Schrödinger and Born-rule forms in the following subsection.
Figure~\ref{fig:FrameFixing} illustrates this frame fixing in the rest frame of the Eulerian observer: the rotation-free boost $T_{\gamma^0}$ maps the timelike leg ${E^\mu}_0$ of a generic vierbein onto the hypersurface normal $n^\mu$, such that the spatial legs ${}^\chi\hspace{-0.22em}{E^\mu}_{\hat{i}}$ become tangent to $\Sigma_t$.

Thus, the total field redefinition is
\begin{align}
    \psi = \left(-g\right)^{-1/4}\sqrt{N} T{_{\gamma^0}}^{-1} \chi,
\end{align}
which implies $\bar{\psi} = \bar{\chi} \left(-g\right)^{-1/4}\sqrt{N} T_{\gamma^0} $ with the new Dirac adjoint $\bar{\chi} = \chi^\dagger \gamma^0$.
As a consequence, the whole redefinition yields the desired form $\sqrt{-g}\bar{\psi}\gamma_t\psi=\bar{\chi}\gamma^0\chi$, relevant for the \sMin forms of the Lagrangian, the inner product, and the quantization rule.
The relation of the field redefinition to an LLT boost is one of our central results, since it shows that, within the proper orthochronous Lorentz group, the rotation-free boost is uniquely fixed, up to local spatial rotations that leave $\gamma^0$ invariant.

\subsection{Redefined Lagrangian, inner product,\\ and quantization rule}
After redefinition, we obtain the Lagrangian
\begin{align}
\label{eq:LagAfterRed}
\begin{split}
    \mathcal{L}=N\bar{\chi}\left[\frac{\ii\hbar c}{2}\gamma_\chi^\mu\lrDalt_\mu- m c^2\right]\chi,
\end{split}
\end{align}
with transformed (boosted) gamma matrices $\gamma_\chi^\mu = T_{\gamma^0} \gamma^\mu {T_{\gamma^0}}^{-1} = {E^\mu}_I {}^{\gamma^0}\hspace{-0.3em}{\Lambda_J}^I \gamma^J={}^\chi\hspace{-0.22em}{E^\mu}_J\gamma^{J}$, with transformed vierbein ${}^\chi\hspace{-0.22em}{E^\mu}_J={E^\mu}_I {}^{\gamma^0}\hspace{-0.3em}{\Lambda_J}^I$, which fulfill the Clifford algebra $\left\lbrace\gamma_\chi^\mu,\gamma_\chi^\nu\right\rbrace=2g^{\mu\nu}$.
Explicitly, we find for the boosted vierbeins ${}^\chi\hspace{-0.22em}{E^\mu}_0= n^\mu$ and ${}^\chi\hspace{-0.22em}{E^\mu}_{\hat{i}}= {E^\nu}_{\hat{i}} h^{\mu\rho}h_{\rho\alpha}\left[ \delta^\alpha_\nu - n_\nu {E^\alpha}_0 / \left( 1 + E^{(n)}_0 \right) \right]$.
Because the redefinition consists of an LLT boost that fixes the vierbein to the Schwinger-time gauge, the boosted vierbeins obey the relations $n_\mu {}^\chi\hspace{-0.22em}{E^\mu}_0 =1$ and $n_\mu {}^\chi\hspace{-0.22em}{E^\mu}_{\hat{i}} =0$, as expected.
The transformed covariant derivative $\lrDalt_\mu$ acts only on $\chi$ and $\bar{\chi}$, where
\begin{align}
\begin{split} 
    \bar{\chi}\gamma_\chi^\mu\lrDalt_\mu\chi=\bar{\chi}\gamma_\chi^\mu\left(\mathfrak{D}_\mu\chi\right)-\left(\mathfrak{D}_\mu\bar{\chi}\right)\gamma_\chi^\mu\chi,
\end{split}
\end{align}
without affecting other prefactors and with
\begin{align}
\begin{split} 
    \mathfrak{D}_\mu\chi=&\left(\partial_\mu+\frac{\ii}{\hbar} q A_\mu+\Gamma^{(\chi)}_\mu\right)\chi,\\
   \mathfrak{D}_\mu\bar{\chi}=&\ \partial_\mu\bar{\chi}-\bar{\chi} \left(\frac{\ii}{\hbar} q A_\mu + \Gamma^{(\chi)}_\mu \right).
\end{split}
\end{align}
Here, \(\mathfrak D_\mu\) is used within the symmetrized bilinear, in which the connection terms associated with the density weight of $\chi$ cancel between the left- and right-acting derivatives.
The derivative includes the boosted spin connection $\Gamma^{(\chi)}_\mu =\omega^{(\chi)}_{\mu IJ}\left[\gamma^I,\gamma^J\right]/8$ with
\begin{align} 
\label{eq:modSpin}
\begin{split}
    \omega^{(\chi)}_{\mu IJ}=g_{\nu\lambda}{{}^\chi\hspace{-0.22em} E^\lambda}_I\big(\partial_\mu {}^\chi\hspace{-0.22em}{E^\nu}_J+\Gamma^\nu_{\mu\sigma}{{}^\chi\hspace{-0.22em}E^\sigma}_J\big).
\end{split}
\end{align}
Although derivatives act on the local rescaling factor, the corresponding terms cancel in the symmetrized kinetic bilinear. 
The remaining change of the spin connection is therefore determined by the local Lorentz boost.
Therefore, only the vierbeins ${E^\mu}_I$ are replaced by the boosted ${}^\chi\hspace{-0.22em}{E^\mu}_I$, also in the spin connection, as expected from an LLT boost.
We observe that the redefined Lagrangian, or the redefined infinitesimal action, respectively, is diffeomorphism- and LLT-independent analogous to the original formulation, due to the invariance of ${E^{\mu}}_I {e_\mu}^{J}$ and the transformation behavior of $\chi$.

In the continuous limit $E^{(n)}=0$, and for finite $\sqrt{-g}/N$, the quantity $K$ does not contain any non-trivial gamma matrices.
Thus, we immediately find in this case $E^{(n)}_0=1$ and $n^\mu={E^\mu}_0$ due to normalization of $n_\mu$ and the common time orientation.
As expected, due to a mere rescaling in this limit, the redefined Lagrangian corresponds to the original formulation with $N$ replacing $\sqrt{-g}$, while keeping the original derivatives.

In general, the connection of the redefined Lagrangian to the inner product can be best observed in its ADM decomposition.
Recall that, by construction, $n_\mu\gamma_\chi^\mu=T_{\gamma^0} n_\mu\gamma^\mu {T_{\gamma^0}}^{-1}=\gamma^0$.
As such, we find in the temporal and spatial ADM decomposition basis the Lagrangian
\begin{align}
\label{eq:LagAfterADM}
\begin{split} 
    \mathcal{L}=\bar{\chi}\left[\frac{\ii\hbar c}{2}\left(\gamma^0\lrDalt{}^{(\tau)}_t+N\gamma_\chi^{(y)i}\lrDalt{}^{(y)}_i\right)-Nmc^2\right]\chi,
\end{split}
\end{align}
with redefined temporal matrix $n_\mu\gamma_\chi^\mu/N=\gamma^0/N$, temporal derivative $\lrDalt{}^{(\tau)}_t=\tau^\nu\lrDalt_\nu$, spatial matrix $\gamma_\chi^{(y)i}=\gamma_\chi^\mu\partial_\mu y^i$, and spatial derivative $\lrDalt{}^{y}_i=\big(\partial^{(y)}_i\!x^\nu\big)\lrDalt_\nu$.
Again, the new gamma matrices in ADM basis fulfill the algebra $\big\lbrace\gamma^0,\gamma^0\big\rbrace/N^2=2/N^2$, $\big\lbrace\gamma^0,\gamma_\chi^{(y)i}\big\rbrace/N=-2N^{(y)i}/N^2$, and $\big\lbrace\gamma_\chi^{(y)i},\gamma_\chi^{(y)j}\big\rbrace=2 \left( h^{(y)ij} + N^{(y)i} N^{(y)j}/N^2 \right)$, in complete analogy to the original gamma matrices in ADM basis.
Due to $\partial_t\chi=c \tau^\nu\partial_\nu\chi$, the Schrödinger form $\mathcal{L}=\ii\hbar\chi^\dagger\lrpartial_t\chi/2-\mathcal{H}$ is restored by the field redefinition, where the temporal and spatial directions are now defined with respect to hypersurfaces of simultaneity. 
The identification of the Hamiltonian density $\mathcal{H}$ therefore proceeds analogously to \sMin quantum field theory, while the non-inertial background appears as an additional contribution to the dynamics, similar to other interactions and forces within the Standard Model.

As expected, the Schrödinger form is directly connected to the redefined inner product
\begin{align}
\label{eq:InnAfterRed}
\begin{split}
    \left(\psi_1,\psi_2\right)=\int_{\Sigma_t} {\mathrm{d}\sigma\bar{\chi}_1\gamma^0\chi_2}=\left(\chi_1,\chi_2\right),
\end{split}
\end{align}
or in $y$ coordinates $\left(\chi_1,\chi_2\right)=\int_{\Sigma_t} {\mathrm{d}^3y\bar{\chi}_1\gamma^0\chi_2}$, which obey the \sMin Born-rule form.
In fact, we obtain the new canonical momentum $\pi_\chi=\ii\hbar \bar{\chi}\gamma^0/2$, having the same algebraic form in $x$ and in $y$ coordinates, and yielding the bilinear form $\left(\chi,\chi\right)=\int{\mathrm{d}^3y\big(\pi_\chi\chi-\bar{\chi}\pi_{\bar{\chi}}\big)/(\ii\hbar)}$, which is consistent with the original representation.

Note that the Schrödinger form of the Lagrangian and the Born-rule form of the inner product persist in $\chi$ representation even under further diffeomorphisms, and both the Lagrangian and the inner product are generally invariant under arbitrary LLTs regardless of the field redefinition. 
However, both Schrödinger and Born-rule form are in general not preserved under a generic additional LLT, since $\gamma^0$ would be replaced by a general $N\gamma_t$ in this case.
In contrast, the world-coordinate frame can still be changed by diffeomorphisms without affecting these structures.
 
Finally, we also transform the Dirac bracket and find the \sMin form 
\begin{equation}
\label{eq:QuantAfterRed}
\left.\left\lbrace\chi_a(x),\bar{\chi}_b(x')\right\rbrace\right|_{\Sigma_t}=\gamma^0_{ab}\delta_{\sigma}(x,x')
\end{equation}
of the quantization rule for two points on the same hypersurface $\Sigma_t$ in arbitrary world coordinates.
In $y$ coordinates, we find the anti-commutator reduces to 
\begin{equation}
    \left.\{\chi_a(t,\vect{y}),\bar\chi_b(t,\vect{y}^\prime)\}\right|_{\Sigma_t}=\gamma^0_{ab}\delta^{(3)}(\vect{y}-\vect{y}^\prime),
\end{equation}
highlighting its equal-time nature explicitly.

\begin{figure*}[t]
    \centering
    \includegraphics{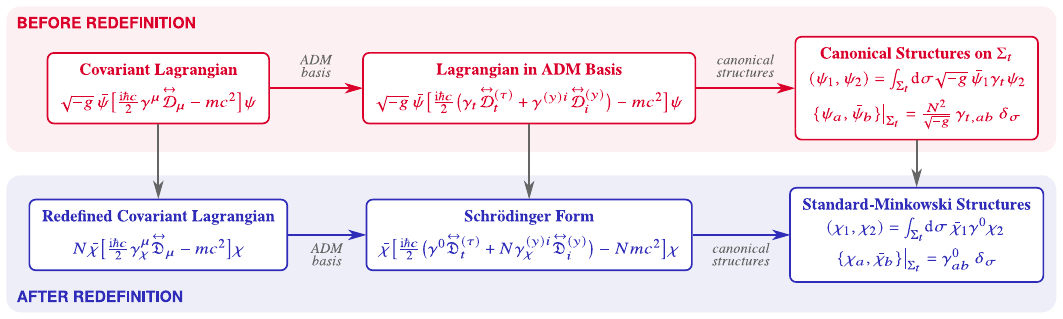}
    \caption{Top row (red): the original Lagrangian, its form in the ADM basis, and the canonical structures it determines on $\Sigma_t$, namely the inner product and the equal-time anti-commutator, are shown. 
    All of them are governed by the background- and foliation-dependent factor $\sqrt{-g}\gamma_t$.
    Bottom row (blue): the same objects after the field redefinition $\psi=(-g)^{-1/4}\sqrt{N}{T_{\gamma^0}}^{-1}\chi$ (vertical arrows) are depicted.
    The rescaling replaces $\sqrt{-g}$ by the lapse function $N$, and the boost $T_{\gamma^0}$ fixes the local Lorentz frame such that $N\gamma_t$ is mapped to $\gamma^0$, see Fig.~\ref{fig:FrameFixing}.
    ADM-decomposing first and redefining afterwards, or redefining the covariant Lagrangian first and ADM-decomposing afterwards, yields the same Schrödinger form, and the value of the inner product is unchanged, $(\psi_1,\psi_2)=(\chi_1,\chi_2)$.
    }
    \label{fig:Routes}
\end{figure*}

Overall, we observe that the redefinition shifts the non-trivial background- and foliation-dependence from the inner product and the quantization rule into an explicit dependency in the Lagrangian and the relation between the original and redefined fields.
As such, before redefinition, there are two pillars for the theory, given by the Lagrangian on the one side, and the inner product and quantization rule on the other. 
After redefinition, the Lagrangian and field redefinition retain the explicit dynamical background dependence, while the inner product and quantization rule assume \sMin form.
Figure~\ref{fig:Routes} summarizes our complete approach: ADM-decomposing the original Lagrangian Eq.~\eqref{eq:LagOriginal} first and redefining the field afterwards, or redefining first and ADM-decomposing afterwards, leads to the same Lagrangian in Schrödinger form from Eq.~\eqref{eq:LagAfterADM}, while the inner product and the quantization rule derived from either Lagrangian are mapped onto each other by the same redefinition.

\section{Applications for redefined theory}
\label{sec:Applications}
To highlight the impact of our formalism, we present three applications in the following.
The first application shows that the usual Newtonian or point-particle potential~\cite{Dodin2010,Poisson2011,Vines2016} arises only after rescaling the fermion field by $\sqrt{-g}/N$.
The second application illustrates the role of fixing the local Lorentz frame, by showing that only this choice highlights the standard spacetime-magnetic or gravito-magnetic~\cite{Jantzen1992,Schafer2004,Costa2014,Ruggiero2023} effect from a modified spin connection, which is essential, \eg{} for the Lense--Thirring effect~\cite{Lense1918,Mashhoon1984}.
Finally, the third application combines both aspects by deriving the non-relativistic momentum operators before and after field redefinition, demonstrating that only the latter assumes the intuitive single-particle form.

\subsection{Newtonian potential from spinor rescaling}
Before rescaling the spinor with $\left(-g\right)^{-1/4}\sqrt{N}$, we find from Eqs.~\eqref{eq:LagOriginal} and \eqref{eq:LagADM} that the rest energy $mc^2$ is multiplied by $\sqrt{-g}$.
However, for a classical point particle at rest with respect to the Eulerian observers, we expect~\cite{Dodin2010,Poisson2011,Vines2016} the well-known lapse-weighted Newtonian potential $N mc^2$.
Indeed, after redefinition, the Lagrangians from Eqs.~\eqref{eq:LagAfterRed} and \eqref{eq:LagAfterADM} include exactly this term for the rest-energy potential.
Thus, the rescaling factor in the redefinition restores the correspondence to the classical Newtonian potential of a point particle.

One prominent example is given by Schwarzschild spacetime, where we find $\sqrt{-g}mc^2=mc^2 r^2\sin\theta$ in Schwarzschild coordinates.
Here, $r$ and $\theta$ are two world coordinates, and $r^2\sin\theta$ is the usual curvilinear functional determinant.
Note that this result is independent of the chosen hypersurface.
However, after redefinition, the lapse-weighted rest energy $Nmc^2=mc^2\sqrt{1-r_\text{S}/r}$, with Schwarzschild radius $r_\text{S}$, is recovered only for the usual constant-foliation-time foliation.
For other choices of the hypersurface and observer, the resulting Newtonian potential generally differs.
This construction is restricted to the exterior region $r>r_\text{S}$, since at the horizon the lapse function vanishes and the redefinition becomes singular, reflecting the breakdown of the static-time foliation.
Hence, this example illustrates that the redefinition (and the constant-foliation-time choice) plays a central role in obtaining the familiar Newtonian potential.

\subsection{Spacetime-magnetic potential from inertial-frame fixing}
The local Lorentz-frame fixing through the redefinition introduces the modified spin connection from Eq.~\eqref{eq:modSpin}, leading to the interaction $\mathcal{L}_\text{sm}=\ii\hbar c N\bar{\chi}\big\lbrace\Gamma^{(\chi)}_\mu,\gamma_\chi^\mu\big\rbrace\chi/2$ in the Lagrangian.
After the linearization $g_{\mu\nu}=\delta^I_\mu\delta^J_\nu\eta_{IJ}+\mathfrak{h}_{\mu\nu}$ around $\eta_{IJ}$, with perturbation $\mathfrak{h}_{\mu\nu}$, we find from the normalization $n_\mu n^\mu=1$ and an expansion around a static-time hypersurface the normal vector $n_\mu\cong\delta^{\bar0}_\mu\left(1+\mathfrak{h}_{\bar0\bar0}/2\right)+\delta^i_\mu\delta n_i$, where $\delta n_i$ is a perturbation of the non-zero hypersurface normal vector.
Consequently, the vierbein\footnote{
Strictly, the vierbein has to be expanded around a non-symmetric tensor perturbation~\cite{Woodard1984} to account for additional LLT degrees of freedom. 
However, for the sake of brevity, we implicitly assumed a special manifold-Lorentz gauge~\cite{Woodard1984,Kostelecky2021} that removes these additional degrees of freedom in the linearized limit, while at the same time still obeying the LLT-fixing conditions from the field redefinition through the Schwinger-gauge choice.
} takes the form ${E^{\mu}}_I\cong\delta^{\mu}_I-\delta^\nu_I\delta^\mu_J\delta^\sigma_K\mathfrak{h}_{\nu\sigma}\eta^{JK}/2$.
Thus, together with the properties of $\gamma^I$, we find
\begin{align}
\begin{split} 
    \mathcal{L}_\text{sm}=\chi^\dagger\frac{\ii\hbar c}{4}\left[\gamma^{\hat{i}},\gamma^{\hat{j}}\right]\delta^i_{\hat{i}}\delta^j_{\hat{j}}\left(\partial_i\delta n_j-\frac{1}{2}\partial_i\mathfrak{h}_{j\bar0}\right)\chi
\end{split}
\end{align}
in the linearized limit, which corresponds in the static-time limit ($\delta n_j=0$) to the spacetime-magnetic or gravito-magnetic effect~\cite{Jantzen1992,Schafer2004,Costa2014,Ruggiero2023} that plays a crucial role for the spin-valued part of the Lense--Thirring frame dragging~\cite{Lense1918,Mashhoon1984}.
For other choices of the hypersurface, this particular spin-connection contribution is modified or even vanishes, while the complete physical effect is redistributed among the remaining terms.
As a result, the explicit appearance of this contribution follows from fixing the local Lorentz frame such that $\gamma_t=\gamma^0/N$, \ie{} fixing Schrödinger and Born-rule forms. 
Without this field redefinition, the corresponding contribution is generally obscured and does not reduce to its familiar static-time expression for choices of the local Lorentz frame.

\subsection{Effects of the field redefinition in the non-relativistic limit}
The fermionic Lagrangian can be organized~\cite{Fischbach1981,Hehl1990,Jentschura2013,Asprea2021,Ito2021,Wang2024} according to 
\begin{align}
\begin{split}
    \mathcal{L}=& \kappa \bar{\psi}\left(-mc^2+ c \frac{\lrE}{2}+ c \frac{\lrO }{2} \right)\psi, 
\end{split}
\end{align}
with diagonal (even) part $\lrE$ and off-diagonal (odd) part $\lrO$, where arrows emphasize that all possibly included covariant derivatives act in a symmetrized way to the left and right as before, and $\kappa$ may depend on world coordinates.
Here, the terms diagonal and off-diagonal refer to inertial $\gamma^I$ matrices in Dirac representation~\cite{Dirac1928,Bjorken1964}, \ie{} $\gamma^{0}$ is diagonal and $\gamma^{\hat{i}}$ off-diagonal.
Next, we block diagonalize the Lagrangian through a local and reversible field redefinition $\psi = \exp\big\{ \rO / \left( 2 mc \right)\big\} \psi^\prime$, akin to a Foldy--Wouthuysen~\cite{Foldy1950,Bjorken1964,Urban1971,Guertin1975,Moss1976,Lin1977,Holstein1997,Pachucki2005,Gardestig2007,Silenko2008,Goncalves2019} transformation.
From this transformation, the new corresponding Dirac adjoint, $\bar{\psi} = \bar{\psi}^\prime \exp\big\{ - \lO/(2mc) \big\}$ follows.
Here, arrows which only point to the right or left indicate non-symmetrized derivatives acting only in this direction.
Note that these non-symmetrized derivatives are defined such that they do not act on the vierbeins standing to the right (left) of them in the right (left) arrow case.
In particular, the single-arrow operators are defined without the minus sign carried by the left-acting derivative within the symmetrized derivative as in Eq.~\eqref{eq:DerSymmBEfore}, \ie{} $\bar{\psi}\lO$ now denotes derivatives of the form $+(\mathcal{D}_\mu\bar{\psi})$, contracted with the respective matrix structure.
The redefinition is constructed such that the new Lagrangian has a modified off-diagonal part $\lrO{}^\prime \propto f\left(\lrE, \lrO \right)/(mc^2)$, where $f$ is a function of both the diagonal and off-diagonal parts, suppressed by the rest energy $mc^2$.
Repeating the transformation once more therefore yields a tree-level diagonalization of the Lagrangian up to order $(mc^2)^{-1}$.
It leads to the non-relativistic Lagrangian
\begin{align}
    \begin{split} \label{eq:NR}
        \mathcal{L}^\text{NR} \cong \kappa \bar{\psi}^\prime \left( -mc^2 + c \frac{\lrE}{2} + \frac{\lO{}^2 -2 \lO\rO + \rO{}^2 }{8m} \right) \psi^\prime.
    \end{split}
\end{align}
As long as we restrict ourselves to momentum states with $c\vect{p} \ll mc^2$, and assume that the relevant background-induced energy scales and spacetime gradients are likewise small compared with the rest-energy scale, the operators within $\lrE$ and $\lrO$ can be treated perturbatively compared to the rest energy in the Lagrangian from Eq.~\eqref{eq:NR}.
This result corresponds to the leading-order non-relativistic limit, in which only the fermionic sector is expanded while the spin-$1$ gauge-field sector remains relativistic.

For the special case of the \sMin Lagrangian in flat spacetime, we find the familiar relations $\lrE = \ii \hbar \gamma^0 \lrD_t$, $\lrO = \ii \hbar \delta^i_{\hat{i}}\gamma^{\hat{i}}\lrD_i$, and $\kappa =1$.
Casting these relations into the above non-relativistic Lagrangian, we obtain Schrödinger form $\mathcal{L}^{\text{NR}}=\ii\hbar\psi^\dagger\lrpartial_t\psi/2-\mathcal{H}^{\text{NR}}$ in the non-relativistic limit, in full analogy to the relativistic \sMin Lagrangian in flat spacetime.
As a consequence, the NR momentum encoded in the off-diagonal element $\lrO$ includes only spatial derivatives, where in these coordinates the indices $i$ label the spatial Cartesian directions.

On curved and non-inertial backgrounds, before redefinition,  the diagonal and off-diagonal parts are
\begin{align}
\begin{split}
    \lrE=&\ \ii\hbar\left(\frac{E^{(n)}_0}{N}\lrD{}^{(\tau)}_t+ E^{i(y)}_0\lrD{}^{(y)}_i\right) \gamma^0 \\
    \lrO=&\  \ii\hbar\left(\frac{E^{(n)}_{\hat{i}}}{N}\lrD{}^{(\tau)}_t+ E^{i(y)}_{\hat{i}}\lrD{}^{(y)}_i\right) \gamma^{\hat{i}}
\end{split}
\end{align}
in ADM decomposition basis, with $\kappa =\sqrt{-g}$.
Here, we defined $E^{i(y)}_0={E^\mu}_0\partial_\mu y^i$ and $E^{i(y)}_{\hat{i}}={E^\mu}_{\hat{i}}\partial_\mu y^i$.
For the respective $\rO{}$ and $\lO{}$ operators, the arrows over the $\lrD{}$ derivatives have to be replaced by $\rD{}$ and $\lD{}$, respectively.
We observe that the corresponding non-relativistic Lagrangian does not assume Schrödinger form due to the factor $\sqrt{-g}E^{(n)}_0/N$ in front of the foliation-time derivative, and we cannot identify an intuitive non-relativistic momentum operator, since the off-diagonal element depends on a foliation-time derivative.

In fact, time derivatives contained in the off-diagonal part $\lrO$ obstruct a direct identification~\cite{Asano2026} of a non-relativistic conjugate momentum consistent with the standard non-relativistic Born rule and complicate a consistent order-by-order non-relativistic expansion.
Indeed, removing these additional time derivatives after the Foldy--Wouthuysen transformation requires~\cite{Asano2026} additional field redefinitions, which mix different orders of the diagonalized Lagrangian and thereby obscure the power counting of the non-relativistic expansion.

In contrast, in our approach we fix the local Lorentz frame in such a way that the diagonal and off-diagonal parts are
\begin{align}
\begin{split}
    \lrE_\chi=&\ \ii\hbar \left( \frac{\lrDalt {}^{(\tau)}_t}{N}+ {}^\chi\hspace{-0.22em}E^{i(y)}_0\lrDalt {}^{(y)}_i\right) \gamma^0 \\
    \lrO_\chi=&\ \ii\hbar {}^\chi\hspace{-0.22em}E^{i(y)}_{\hat{i}}\lrDalt {}^{(y)}_i\gamma^{\hat{i}},
\end{split}
\end{align}
and we rescale the spinor\footnote{Obtaining an inner product and a bilinear form that both assume \sMin form in the non-relativistic limit may require performing the non-relativistic expansion before the spinor rescaling. A systematic analysis of this ordering lies beyond the scope of the present work.} such that $\kappa=N$, ${}^\chi\hspace{-0.22em}E^{i(y)}_0={}^\chi\hspace{-0.22em}{E^\mu}_0\partial_\mu y^i$, and ${}^\chi\hspace{-0.22em}E^{i(y)}_{\hat{i}}={}^\chi\hspace{-0.22em}{E^\mu}_{\hat{i}}\partial_\mu y^i$. 
Again, for the respective $\rO{}$ and $\lO{}$ operators, the arrows over the $\lrDalt{}$ derivatives have to be replaced by $\rDalt{}$ and $\lDalt{}$, respectively.
Upon inserting the diagonal and off-diagonal parts, as well as $\kappa$ into the non-relativistic Lagrangian, we find that the redefinition restores Schrödinger form also in the non-relativistic limit.
Moreover, the resulting non-relativistic momentum operator is defined solely through spatial derivatives, recovering the form expected for single-particle quantum mechanics upon projecting on a one-particle subspace~\cite{Birrell1982,Wald2001,Parker2009}. 
This result naturally follows from the connection between the field redefinition and the Schwinger time gauge~\cite{Schwinger1963} mentioned in Sec.~\ref{sec:RedefFieldTheory} and illustrated in Fig.~\ref{fig:FrameFixing}.
In this gauge, the new projected vierbein satisfies $n_\mu{}^\chi\hspace{-0.22em}{E^\mu}_0 = 1$ and $n_\mu{}^\chi\hspace{-0.22em}{E^\mu}_{\hat{i}} =0$, which is commonly used~\cite{Leclerc2006} for obtaining a proper Hamiltonian description in non-inertial backgrounds.
Even more, our formalism and the results of the previous section~\ref{sec:RedefFieldTheory} show that this gauge, fixing the local Lorentz frame, is not merely a convenient choice among others for deriving the non-relativistic limit of the Lagrangian with purely spatial momentum.
Rather, it is \textit{uniquely} selected within the continuous branch considered here by the requirement of recovering this standard non-relativistic structure.

\section{Discussion}
\label{sec:Conclusion}
Quantum field theories on curved or non-inertial backgrounds have been formulated~\cite{Birrell1982,Parker2009,Buchbinder2021} for fermions as well as gauge fields in a variety of settings, including general backgrounds and perturbative linearizations.
While quantization procedures~\cite{Leclerc2007,Nelson1978} and ambiguities in the choice of the vacuum~\cite{Fulling1973,Davies1975,Unruh1976,Hawking1974,Wald2001,Crispino2008} have been studied, explicit Hamiltonian formulations and particle interpretations frequently specialize to stationary backgrounds or particular choices for foliations.
In particular, such field-theoretical studies generally do not employ a field redefinition that simultaneously maps the fermionic Lagrangian, inner product, and canonical quantization rule to their respective \sMin forms.

For single-particle quantum mechanics on non-inertial backgrounds, often special cases involving particular choices of hypersurfaces and backgrounds have been considered, both for scalar~\cite{Lammerzahl1995,Marzlin1995,Schwartz2019,Exirifard2022} and for spinor~\cite{Fischbach1981,Hehl1990,Jentschura2013,Asprea2021,Ito2021,Alibabaei2023,Wang2024} theories.
In these limits, also corresponding wave-function redefinitions~\cite{Fischbach1981,Lammerzahl1995,Jentschura2013,Asprea2021,Perche2021} have been studied.
However, to our knowledge, these considerations have not been combined within a single field-theoretical framework for general spacelike foliations and curved or non-inertial backgrounds, while simultaneously establishing a connection between the \sMin form of the (single-particle) inner product and the quantization rule to the Schrödinger form of the Lagrangian. 

In this work, we present a reformulated field-theoretical framework for fermions coupled to general spin-$1$ gauge fields on curved and non-inertial backgrounds with a suitable spacelike foliation, such that the inner product from Eq.~\eqref{eq:InnAfterRed} and the quantization rule from Eq.~\eqref{eq:QuantAfterRed} take their respective \sMin forms.
In this context, we show that the resulting Lagrangian from Eq.~\eqref{eq:LagAfterADM} assumes Schrödinger form in a generalized temporal and spatial ADM decomposition.
Although the \sMin forms are recovered, the theory still features general curved or non-inertial backgrounds and suitable hypersurfaces of simultaneity.
In our approach, the manifestly covariant form of the Lagrangian is replaced by a foliation-dependent representation in which LLT-covariance is no longer manifest, while we demonstrate that the underlying symmetries are preserved through the induced transformation laws of the redefined fields.
We find that the redefinition consists of a rescaling, which converts the factor $\sqrt{-g}$ multiplying the fermionic bilinear into the lapse factor $N$, and a transformation that fixes the local Lorentz frame such that the ADM temporal matrix $\gamma_t$ is mapped to $\gamma^0/N$.
Figure~\ref{fig:Routes} summarizes how the individual structures of the theory change under both parts of the redefinition and shows that ADM decomposition and redefinition can be performed in either order.
We derive the latter transformation as a general field redefinition, naturally arising from the $\gamma_t\rightarrow\gamma^0/N$ requirement, but also associate it with a specific LLT.
We show that this choice is \textit{uniquely} selected by the requirement of recovering the desired \sMin forms.
This approach differs from previous works, which focused on more specialized settings, and where either the connection to LLTs was not established~\cite{Gorbatenko2010,Gorbatenko2011} or where an LLT was directly presumed~\cite{Beneke2023}.
Moreover, by making use of the symmetrized derivative, we render the Hermitian structure of the fermionic kinetic term manifest~\cite{Leclerc2006} and avoid spurious pure-derivative terms involving the background~\cite{Huang2009}.
We apply our results to highlight that our field redefinition converts the coordinate-density factor $\sqrt{-g}$ multiplying the mass into the lapse-weighted rest-energy term $Nmc^2$, in agreement~\cite{Dodin2010,Poisson2011,Vines2016} with a classical point particle at rest with respect to the Eulerian observers.
Furthermore, we show that fixing the local Lorentz frame by mapping $\gamma_t$ to $\gamma^0/N$ makes the familiar spacetime-magnetic or gravito-magnetic effect~\cite{Jantzen1992,Schafer2004,Costa2014,Ruggiero2023} explicit.
Finally, we demonstrate that both components of the redefinition play complementary roles in obtaining a well-defined non-relativistic formulation with a momentum operator containing only spatial derivatives.
These applications highlight that the redefinition makes several established effects and methods more transparent or even useful in an intuitive way by collecting the relevant background dependence into a \sMin formulation with a fixed timelike direction of the local Lorentz frame.

Our results provide a starting point for studying quantum-field-theoretical effects in strong curved or non-inertial backgrounds, as well as in the linearized limit relevant for terrestrial experiments.
Because the \sMin forms of the inner product and quantization rule are recovered, our approach may facilitate computations for particle-physics observables in the presence of non-inertial effects.
Examples include precision measurements of the anomalous magnetic moment~\cite{Morishima2018,Ulbricht2019} or the Lamb shift~\cite{Audretsch1995,Alvarez1996,Zhou2012,Arya2023}.
Moreover, the redefined formulation could be applied to scenarios~\cite{Bose2017,Marletto2017,Rossi2025} where quantum matter serves as a source mass for gravity, with possible applications to the construction and characterization of Hadamard states~\cite{Radzikowski1996,Hollands2001}, semiclassical gravity~\cite{Birrell1982,Wald2001,Hu2020}, stochastic gravity~\cite{Verdaguer2007,Hu2008}, or quantum theories of gravity~\cite{Kiefer1991,Bomstad2006,Maniccia2023,Oppenheim2023}.
The redefined theory could also form the basis for the fermionic sector of an effective field theory~\cite{Paz2015,Berwein2019,Burgess2020}, such as the Standard Model effective field theory~\cite{Isidori2024}, on curved or non-inertial backgrounds~\cite{Padmanabhan2011,Ruhdorfer2020,Kostelecky2021,Beneke2023}, while maintaining the \sMin forms of the inner product and quantization rule for fermions.
In addition, the non-relativistic expansion can be extended to higher orders~\cite{Asano2026} to describe low-energy experiments where the \sMin forms of the fermionic inner product and quantization rule allow for a transparent interpretation of interactions and observables, as shown by our applications.
This limit may also facilitate the study of composite particles~\cite{Sonnleitner2018,Schwartz2019_2,Asano2024} and bound-state calculations~\cite{Caswell1986,Pineda1998,Escobedo2008}, as well as a generalized quantum-field-theoretical mass defect~\cite{Zych2011,Sonnleitner2018,Schwartz2019_2,Asano2024}.
Lastly, one can investigate whether analogous redefinitions can simplify the hypersurface inner products and corresponding structures~\cite{Peskin1995,Weinberg1995,Wald2001} of gauge fields while accounting for their gauge constraints.

\section*{Acknowledgments}
We are grateful to W. P. Schleich for his continuing support.
We also thank C. Niehof for valuable input, as well as the QUANTUS team for fruitful and interesting discussions.
The authors also acknowledge contributions in the form of discussions from the Terrestrial Very-Long-Baseline Atom Interferometry (TVLBAI) proto-collaboration.
The QUANTUS project is supported by the German Space Agency at the German Aerospace Center (Deutsche Raumfahrtagentur im Deutschen Zentrum f\"ur Luft- und Raumfahrt, DLR) with funds provided by the Federal Ministry for Economic Affairs and Climate Action (Bundesministerium f\"ur Wirtschaft und Klimaschutz, BMWK) due to an enactment of the German Bundestag under Grant No. 50WM2450D and No. 50WM2450E (QUANTUS-VI).

\section*{Tools and Resources}
The authors directed generative artificial-intelligence tools, namely OpenAI ChatGPT (models 5.5 and 5.6) and Anthropic Claude (Opus 4.8 to 5.0 as well as Fable 5.0 and 5.1), the latter in part through the Claude Code interface, to (i) independently re-derive the analytic results of this article and check them symbolically, (ii) draft and debug the SymPy verification scripts, and (iii) suggest improvements of presentation, including the design of the figures, and copyediting of language. 
All output was reviewed and verified by the authors, who take full responsibility for the entire content of this article.

All figures were drawn by the authors in Inkscape.

\section*{Author declarations}
\subsection*{Conflict of interest}

\noindent The authors have no conflicts to disclose.

\subsection*{Data availability}
\noindent This article contains no data sets beyond the analytic results shown in the text; the exact symbolic verification package that reproduces all derivation steps is openly available on GitHub~\cite{VerificationLedger2026} and will be archived on Zenodo upon publication.

\bibliography{Literatur}

\end{document}